\pdfoutput=1
\documentclass[%
reprint,
superscriptaddress,
nofootinbib,
 amsmath,amssymb,
 aps,
 prx,
 longbibliography,
floatfix,
]{revtex4-1}

\usepackage{graphicx}
\usepackage{dcolumn}
\usepackage{bm}
\usepackage{appendix}
\usepackage{hyperref}
\usepackage{tabularx}
\usepackage{graphicx}
\usepackage{amsmath}
\usepackage{blkarray, bigstrut}
\usepackage{amsthm}
\usepackage{algorithm}
\usepackage{algpseudocode}
\usepackage{nicematrix}
\usepackage{physics}

\theoremstyle{definition}

\usepackage{makecell}
\setcellgapes{5pt}
\usepackage[OT2,T1]{fontenc}
\DeclareSymbolFont{cyrletters}{OT2}{wncyr}{m}{n}
\DeclareMathSymbol{\Sha}{\mathalpha}{cyrletters}{"58}

\usepackage{mathtools}

\usepackage{array,amssymb,booktabs}
\newcolumntype{C}{>{$}c<{$}}
\AtBeginDocument{
    \heavyrulewidth=.08em
    \lightrulewidth=.05em
    \cmidrulewidth=.03em
    \belowrulesep=.65ex
    \belowbottomsep=0pt
    \aboverulesep=.4ex
    \abovetopsep=0pt
    \cmidrulesep=\doublerulesep
    \cmidrulekern=.5em
    \defaultaddspace=.5em
}

\begin{document}

\preprint{APS/123-QED}

\title{Experimentally realized \textit{in situ} backpropagation for deep learning in nanophotonic neural networks}

\author{Sunil Pai}
\email{sunilpai@stanford.edu}
\affiliation{Department of Electrical Engineering, Stanford University, Stanford, CA 94305, USA}
\author{Zhanghao Sun}
\affiliation{Department of Electrical Engineering, Stanford University, Stanford, CA 94305, USA}
\author{Tyler W. Hughes}
\affiliation{Department of Electrical Engineering, Stanford University, Stanford, CA 94305, USA}
\affiliation{now at Flexcompute Inc., Belmont, MA, USA}
\author{Taewon Park}
\affiliation{Department of Electrical Engineering, Stanford University, Stanford, CA 94305, USA}
\author{Ben Bartlett}
\affiliation{Department of Applied Physics, Stanford University, Stanford, CA 94305, USA}
\author{Ian A. D. Williamson}
\affiliation{Department of Electrical Engineering, Stanford University, Stanford, CA 94305, USA}
\affiliation{now at X Development LLC, Mountain View, CA USA.}
\author{Momchil Minkov}
\affiliation{Department of Electrical Engineering, Stanford University, Stanford, CA 94305, USA}
\affiliation{now at Flexcompute Inc., Belmont, MA, USA}
\author{Maziyar Milanizadeh}
\affiliation{Politecnico di Milano, Milan, Italy}
\author{Nathnael Abebe}
\affiliation{Department of Electrical Engineering, Stanford University, Stanford, CA 94305, USA}
\author{Francesco Morichetti}
\affiliation{Politecnico di Milano, Milan, Italy}
\author{Andrea Melloni}
\affiliation{Politecnico di Milano, Milan, Italy}
\author{Shanhui Fan}
\affiliation{Department of Electrical Engineering, Stanford University, Stanford, CA 94305, USA}
\author{Olav Solgaard}
\affiliation{Department of Electrical Engineering, Stanford University, Stanford, CA 94305, USA}
\author{David A.B. Miller}
\affiliation{Department of Electrical Engineering, Stanford University, Stanford, CA 94305, USA}

\begin{abstract}
Neural networks are widely deployed models across many scientific disciplines and commercial endeavors ranging from edge computing and sensing to large-scale signal processing in data centers. The most efficient and well-entrenched method to train such networks is backpropagation, or reverse-mode automatic differentiation.
To counter an exponentially increasing energy budget in the artificial intelligence computing sector, there has been recent interest in analog implementations of neural networks, specifically nanophotonic optical neural networks for which no analog backpropagation demonstration exists.
We design mass-manufacturable silicon photonic neural networks that alternately cascade our custom designed ``photonic mesh'' accelerator with digital nonlinearities to output the result of arbitrary matrix multiplication of the input signal. 
These photonic meshes are parametrized by reconfigurable physical voltages that tune the interference of optically encoded input data propagating through integrated Mach-Zehnder interferometer networks.
Here, using our packaged photonic chip, we demonstrate \textit{in situ} backpropagation for the first time to solve classification tasks and evaluate a new protocol to keep the entire gradient measurement and update of physical device voltages in the analog domain, improving on past theoretical proposals.
This \textit{in situ} method is made possible by introducing three changes to typical photonic meshes: (1) measurements at optical ``grating tap'' monitors, (2) bidirectional optical signal propagation automated by fiber switch, and (3) universal generation and readout of optical amplitude and phase.
After training, our classification achieves accuracies similar to digital equivalents even in the presence of systematic error. Our findings suggest a new training paradigm for photonics-accelerated artificial intelligence based entirely on a physical analog of the popular backpropagation technique.
\end{abstract}

\pacs{85.40.Bh}
\keywords{universal linear optics, machine learning}
\maketitle

Neural networks (NNs) are intelligent computational graph-based models that are ubiquitous in scientific data analysis and commercial artificial intelligence (AI) applications such as self-driving cars and speech recognition software. Through deep learning, NN models are dynamically ``trained'' on input image, audio or language data to automatically make decisions (``inference'') for complex signal processing powering much of today's modern technology. Due to increasing demand, these models require an ever-increasing computational energy budget, which has recently been estimated to double every 3 to 4 months, according to OpenAI \cite{Amodei2018AICompute}. An increasingly large reservoir of available data and adoption of AI in modern technology necessitates an energy-efficient solution for training of NNs. 

In this paper, we experimentally demonstrate the first (to our knowledge) optical implementation of \textit{backpropagation}, the most widely used and accepted method of training NNs \cite{Linnainmaa1976TaylorError, Rumelhart1986LearningErrors}, on a scalable foundry-manufactured device.
(A minimal bulk optical demonstration has been previously explored  \cite{Cruz-Cabrera2000ReinforcementEffects}.)
Specifically, backpropagation consists of backward propagating model errors computed for training data through the NN graph to determine updating gradients on each element in the graph. 
Importantly, this can be physically implemented in linear optical devices by simply sending light-encoded errors backwards through photonic devices and performing optical measurements \cite{Hughes2018TrainingMeasurement}, which is a faster and more efficient calculation than a digital implementation.
Our demonstration of this new physics-based backpropagation algorithm, and analysis of systematic error in gradient calculations used in backpropagation, could help ultimately offer new, possibly energy-efficient strategies to teach modern AI to make intelligent decisions on mass manufacturable silicon photonics hardware using efficient model-free training \cite{Hughes2018TrainingMeasurement, Pai2020ParallelNetwork}. 

\begin{figure*}
    \centering
    \includegraphics[width=\textwidth]{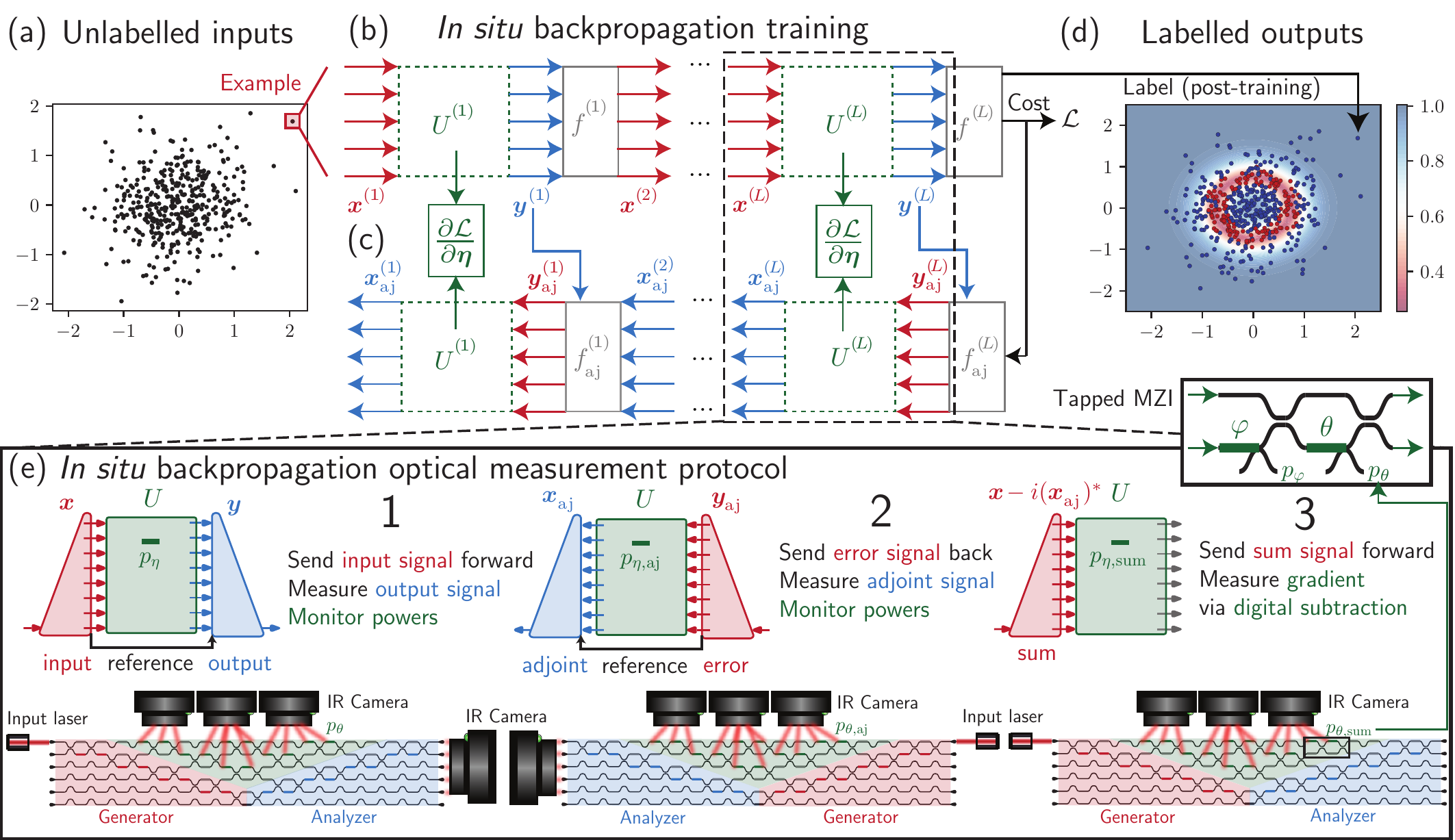}
    \caption{(a) Example machine learning problem: an unlabelled 2D set of inputs that are formatted to be input into a photonic neural network. (b, c) \textit{In situ} backpropagation training of an $L$ photonic neural network for (b) the forward direction and (c) the backward direction showing the procedure for calculating gradient updates for phase shifts. (d) An inference task implemented on the actual chip results in good agreement between the chip-labelled points and the ideal implemented ring classification boundary (resulting from the ideal model) and a 90\% classification accuracy. (e) We depict our proposed architecture and the three steps of \textit{in situ} (analog) backpropagation, consisting of a $6 \times 6$ mesh implementing coherent $4 \times 4$ forward and inverse unitary matrix-vector products using a reference arm. We depict the (1) forward (2) backward (3) sum steps of \textit{in situ} photonic backpropagation. Arbitrary input setting and complete amplitude and phase output measurement are enabled in both directions using the reciprocity and symmetries of our architecture. All powers throughout the mesh are monitored using the tapped MZI shown in the inset for each step, allowing for digital subtraction to compute the gradient \cite{Hughes2018TrainingMeasurement}. These power measurements performed at phase shifts are indicated by green horizontal bars.}
    \label{fig:backprop}
\end{figure*}

As a physical platform for our backpropagation demonstration, we explore programmable nanophotonic devices called ``photonic meshes'' for accelerating matrix multiplication \cite{Shen2017DeepCircuits, Miller2013Self-configuringInvited}.
Photonic meshes shown in Fig. \ref{fig:backprop} are silicon-based low-cost, commercially scalable $N\times N$ port photonic integrated circuits (PICs) consisting of Mach-Zehnder interferometers (MZIs) and programmable optical phase shifts.
These PICs are capable of representing matrix-vector multiplication (MVM) through only the propagation of guided monochromatic light (1560 nm in our demonstration) through $N$ silicon waveguide ``wires'' clad by silicon oxide \cite{Miller2013Self-configuringInvited, Annoni2017UnscramblingModes, Shen2017DeepCircuits}.
Each waveguide can support a single optical mode which has two degrees of freedom: amplitude and phase, yielding a complex $N$-dimensional vector $\bm{x}$ at the input of the system.
Programmable phase shift settings physically modulate the propagation speed (and relative phases) of the wave over segments of silicon wire to affect how the $N$ propagating modes constructively or destructively interfere in each interferometer.
The energy efficiency of these devices has been estimated to be up to two orders of magnitude higher than current state-of-the-art electronic application-specific integrated circuits (ASICs) in AI \cite{Nahmias2020PhotonicNetworks}.

Assuming no light is lost in the ideal photonic circuit, the mesh can be programmed to transform optical inputs using an arbitrary programmable unitary MVM $\bm{y} = U \bm{x}$  \cite{Hurwitz1897UberIntegration, Reck1994ExperimentalOperator, Miller2013Self-configuringInvited}. 
The matrix $U$ is parametrized by the programmable phase shifters on the device and transforms inputs $\bm{x}$ propagating through the device to output modes $\bm{y}$.
The programmed phase shifts on the device define the matrix $U$, and the $N$ output mode amplitude and phase measurements $\bm{y}$ represent the solution to this optical computation.
This fundamental mathematical operation enables meshes to be widely employed in various analog signal processing applications \citenum{Bogaerts2020ProgrammableCircuits} such as telecommunications \cite{Annoni2017UnscramblingModes}, quantum computing \cite{Carolan2015UniversalOptics}, sensing, and machine learning \cite{Shen2017DeepCircuits}, the last of which we explore experimentally in this work via our backpropagation demonstration.

To form what we call a ``hybrid'' digital-photonic NN (PNN), we alternately cascade photonic meshes and digital nonlinear functions \cite{Bogaerts2020ProgrammableCircuits, Harris2018LinearProcessors}, which ultimately forms a composite function and model capable of complex decision making. While performing inference or backpropagation, the hybrid PNN performs time-and energy-efficient MVM, converts photonic mesh output signals to the digital domain, applies nonlinearities, and then converts the data back to optical domain for MVM in the next layer. Hybrid PNNs offer more versatility over fully analog PNNs in the near term due to flexible manipulation of signals in the digital domain easing implementations for recurrent and convolutional neural networks. As a result, hybrid PNNs have been demonstrated to provide a reliable low-latency and energy-efficient analog optical solution for inference, recently in circuit sizes of up to $64 \times 64$ in commercial settings \cite{LightmatterCompany.}. Despite this success in PNN-based inference, on-device backpropagation training of PNNs has not been demonstrated, due to significantly higher experimental complexity compared to the inference procedure.

In this paper, we address this gap by experimentally demonstrating \textit{in situ} backpropagation in a hybrid PNN architecture. 
First, we propose a novel and energy efficient implementation of the technique for measuring phase shifter updates entirely in the optoelectronic (analog) domain. 
Second, we experimentally validate training a backpropagation-enabled, foundry-manufactured photonic circuit using a custom optical rig setup on a multilayer neural network. 
Our demonstration solves machine learning tasks on this photonic hardware using this optically-accelerated backpropagation with similar accuracy compared to a conventional digital implementation, adding new capabilities beyond existing inference or \textit{in silico} learning demonstrations \cite{Shen2017DeepCircuits, Wright2022DeepBackpropagation, Spall2022HybridNetworks}. 
Our findings ultimately pave the way for a new class of approaches for energy-efficient analog training of neural networks and optical devices more broadly.

\begin{figure*}
    \centering
    \includegraphics[width=\textwidth]{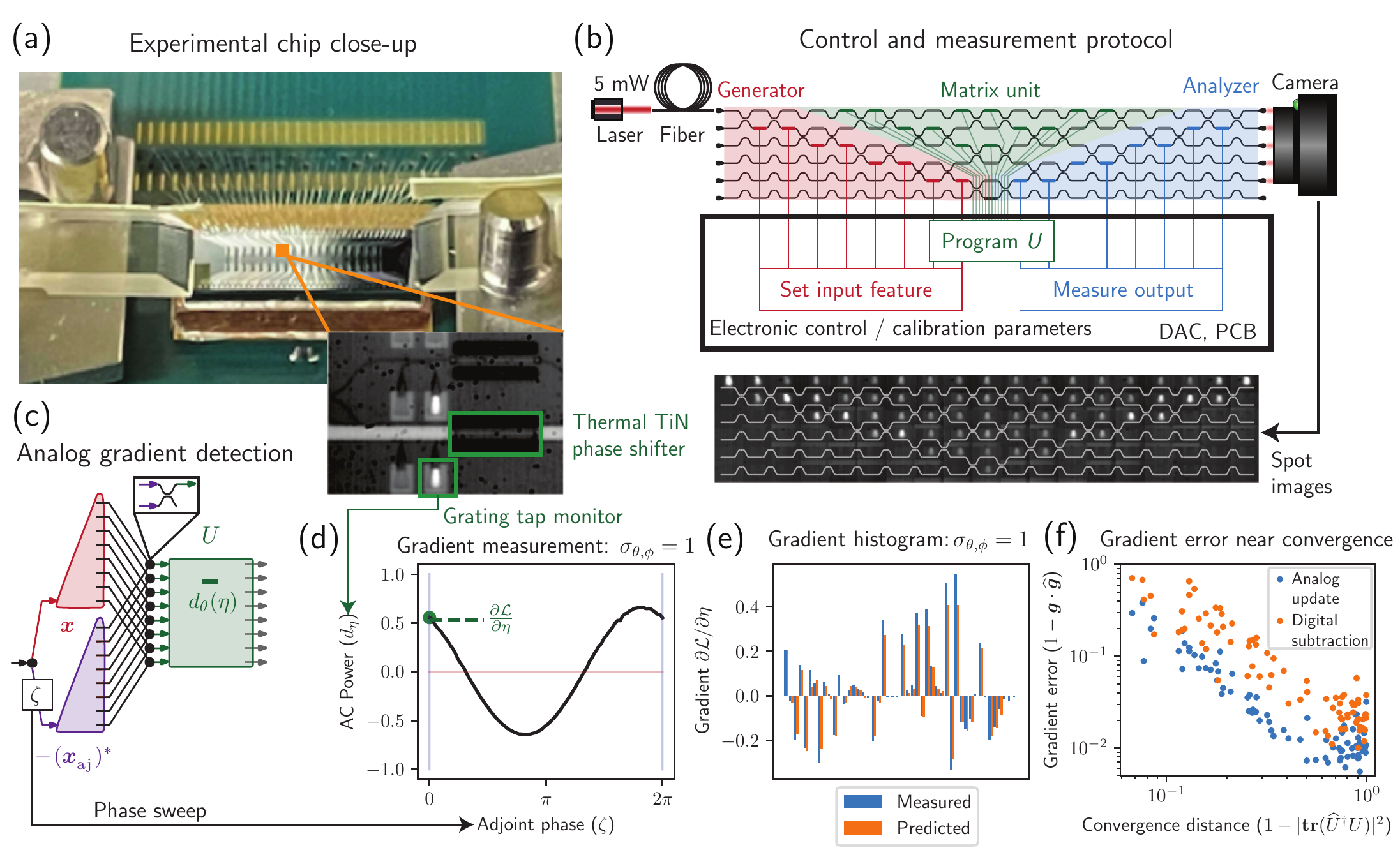}
    \caption{(a) Image of the packaged photonic chip wirebonded to a custom PCB with fiber array for laser input and a camera overhead for imaging the chip. Zooming in reveals the core control-and-measurement unit of the chip, enabling power measurement using 3\% grating tap monitors and a thermal phase shifter nearby. (b) The DAC control is used to set up inputs and perform coherent detection of the outputs, and the IR camera over the chip can be used to image all waveguide segments throughout the chip via grating tap monitors, which we use for backpropagation. (c) Analog gradient detection could be used to measure the gradient by introducing a summing interference circuit (not implemented on the chip in (b)) between the input and adjoint fields. (d) The adjoint phase $\zeta$ can be swept from 0 to $2\pi$ to perform an all-analog gradient measurement, and, ultimately in principle, update phase shifters with an optoelectronic scheme. (e) Gradients measured using our analog scheme yield approximately correct gradients when the implemented mesh is perturbed from the optimal (target) unitary $U = \mathrm{DFT}(4)$ with phase error $\sigma_{\theta, \phi} = 1$. (f) The average normalized gradient error (averaged over 20 instances of random implemented $\widehat{U}$) decreases with distance of the device implementation $\hat{U}(\bm{\eta})$ from the optimal $U = \mathrm{DFT}(4)$. This ``distance'' is represented in terms of the fidelity error $1 - |\mathbf{tr}(\widehat{U}^\dagger U)|^2$.}
    \label{fig:backpropgradient}
\end{figure*}

\section*{Photonic neural networks}

At a high level, a neural network is able to transform data into useful decisions or interpretations, an example of which is shown in Fig. \ref{fig:backprop}(a) and (d) where we label points in 2D based on their location within or outside of a ring. This problem (and in principle many more complex problems like audio signal processing \cite{Shen2017DeepCircuits}) can be solved in the optical domain using photonic neural networks (PNNs) as shown in Fig \ref{fig:backprop}(a)-(d). To solve the problem, we design and evaluate a hybrid digital-optical deep PNN architecture parameterized by trainable programmable phase shifts $\bm{\eta} \in [0, 2\pi)^D$, where $D$ represents the total number of phase shifting elements across all layers in the overall PNN.

Using a combination of photonic hardware and software implementations, our hybrid PNN can solve nontrivial tasks using alternating sequences of analog linear optical MVM operations $U^{(\ell)}(\bm{\eta}^{(\ell)})$ and digital nonlinear transformations $\bm{f}^{(\ell)}$ where $\ell$ denotes the neural network layer and we assume a total of $L$ layers. For example, after $L = 3$ neural network layers for $N = 4$, our photonic neural network is capable of transforming the boundary function used to separate the labelled points in Fig. \ref{fig:backprop}(d). To realize this implementation in a mathematical model, the following sequence of functions transforms the data, proceeding in a ``feedforward'' manner through the layers of the network:
\begin{equation}\label{eqn:onn}
    \begin{aligned}
        \bm{y}^{(\ell)} &= U^{(\ell)} \bm{x}^{(\ell)}\\
        \bm{x}^{(\ell + 1)} &= f^{(\ell)}(\bm{y}^{(\ell)}).
    \end{aligned}
\end{equation}
The inputs $\bm{x} = \bm{x}^{(1)}$ to the overall system are forward-propagated to the final layer (layer $L$), outputting $\widehat{\bm{z}} := \bm{x}^{(L + 1)}$. This forward propagation and resulting output measurement of data sent through this network is called ``inference'' and is depicted in Fig. \ref{fig:backprop}(a, b, d). The model cost or error function is represented by $\mathcal{L}(\bm{x}, \bm{z}) = c(\widehat{\bm{z}}(\bm{x}), \bm{z})$ for a given set of ground truth labels $\bm{z}$, where $c$ is any cost function representing the error between $\widehat{\bm{z}}$ and $\bm{z}$. We refer to the input, label pair of training data $(\bm{x}, \bm{z})$ as a ``training example.'' Backpropagation and other gradient-based training approaches seek to update parameters $\bm{\eta}$ based on the vector gradient $\frac{\partial \mathcal{L}}{\partial \bm{\eta}} \in \mathbb{R}^D$ evaluated for a given training example (or averaged over a batch of training examples). We now explain how we implement both the inference and backpropagation training calculations directly on our core photonic neural network.

\section*{Backpropagation demonstration}

For practical demonstration purposes, our multilayer PNN is completely controlled by a single a photonic mesh (Note that in practice, energy-efficient photonic NNs are controlled by separate photonic meshes of MZIs for each linear layer). Each MZI unit is controlled by an electronic control unit that applies voltages to set various phase shifts on the device packaged on a thermally controlled assembly as shown in Fig. \ref{fig:backpropgradient}(a, b). These phase shifts are placed at the input external arm of the MZI ($\phi$, controlled by voltage $v_\phi$) and in the internal arm of the MZI ($\theta$ controlled by voltage $v_\theta$); this ultimately controls the propagation pattern of the light through the chip, enabling arbitrary unitary matrix multiplication. In our chip specifically, we embed an arbitrary $4 \times 4$ unitary matrix multiply in a $6 \times 6$ triangular network of MZIs. This configuration incorporates two $1 \times 5$ photonic meshes on either end of the $4 \times 4$ ``Matrix unit'' (shown in green) capable of sending any input vector $\bm{x}$ and measuring any output vector $\bm{y}$ from Eq. \ref{eqn:onn}. These calibrated optical I/O circuits are referred to as ``Generator'' and ``Analyzer'' circuits are shown in red and blue respectively in Figs. \ref{fig:backprop}(e) and \ref{fig:backpropgradient}(b). A more complete discussion of how an MVM operation is achieved using our architecture is provided in the Methods, and similar approaches have been attempted for complex-valued photonic neural network architectures \cite{Zhang2021AnNetwork}. Note that for the input generation and output measurements, we need to calibrate the voltage mappings $\theta(v_\theta), \phi(v_\phi)$ (equivalently for the output measurement, $v_\theta(\theta), v_\phi(\phi)$), which is discussed in detail in Ref. \citenum{Miller2020AnalyzingNetworks}. This is a standard calibration protocol  \cite{Shen2017DeepCircuits, Prabhu2020AcceleratingCircuits, Miller2015PerfectComponents} discussed at length in the Appendix and required for accurate operation of the chip.

Our core contribution in this paper, shown in Fig. \ref{fig:backprop}(e), is to devise and test a photonic mesh matrix accelerator architecture that experimentally implements backpropagation as proposed in Ref. \citenum{Hughes2018TrainingMeasurement} within a hybrid digital-analog model implementing the most expensive operations using universal linear optics. Our backpropagation-enabled architecture differs from previously-proposed photonic mesh architectures in three ways:
\begin{enumerate}
    \item We enable ``bidirectional light propagation,'' the ability to send and measure light propagating left-to-right or right-to-left through the circuit (as depicted in Fig. \ref{fig:backprop}(e)).
    \item We implement ``global monitoring,'' the ability to measure optical power at any waveguide segment in the circuit using 3\% grating taps (shown in the inset of Fig. \ref{fig:backprop}(e) and Fig. \ref{fig:backpropgradient}(a, b)). In our proof-of-concept setup, we use an IR camera mounted on an automated stage to image these taps throughout the chip.
    \item We implement both amplitude and phase detection (improving on past approaches \cite{Zhang2021AnNetwork}) using a self-configuring programmable Matrix unit layer \cite{Miller2017SettingMethod, Miller2020AnalyzingNetworks} on both the red and blue Generator and Analyzer subcircuits of Fig. \ref{fig:backprop}(e) and Fig. \ref{fig:backpropgradient}(b), which by symmetry works for sending and measuring light that propagates forward or backward through the mesh.
\end{enumerate}

These improvements on an already versatile photonic hardware architecture enable backpropagation-based machine learning implemented entirely using optical measurement to optimize programmable phase shifters in PNNs. As shown in Fig. \ref{fig:backprop}(e), all three steps of backpropagation \citenum{Hughes2018TrainingMeasurement} require monitoring of optical powers at each phase shifter and measurement of complex field outputs at the left and right sides of the mesh.
Furthermore, the bidirectionality of the monitoring and optical I/O is required to switch between forward and backward propagation of signals required for \textit{in situ} backpropagation to be experimentally realized.
Equipped with these additional elements, our protocol can be implemented on any feedforward photonic circuit \cite{Pai2020ParallelNetwork} with the requisite Analyzer and Generator circuitry, though we use a triangular mesh in this work to enable the chip to be used in other applications \cite{Miller2013EstablishingAutomatically}.

Here we give a quick summary of the procedure (fully described in the Appendix). For each layer $\ell$ and training example pair, a ``forward inference'' signal $\bm{x}^{(\ell)}$ is sent forward and a corresponding ``backward adjoint'' signal $\bm{x}_{\mathrm{aj}}^{(\ell)}$ is sent backward through a mesh implementing $U^{(\ell)}$. The backward pass is in a sense a mirror image of the forward pass (error signal is sent from final layer to input layer) which algorithmically computes an efficient ``reverse mode'' chain rule calculation. The final step sends what we call a ``sum'' vector $\bm{x}^{(\ell)} - i (\bm{x}^{(\ell)}_{\mathrm{aj}})^*$. Previously \cite{Hughes2018TrainingMeasurement}, it was shown that global monitoring in all three steps enables us to calculate the gradient by subtracting the backward and forward measurements from sum measurements in the digital domain, in what we call an ``optical vector-Jacobian product (VJP)'' (Appendix).

\section*{Analog update}

Going beyond an experimental implementation of the theoretical proposal of Ref. \citenum{Hughes2018TrainingMeasurement}, we additionally explore a more energy-efficient fully analog gradient measurement update for the final step that avoids the digital subtraction update. The key difference is in the final ``sum'' step where we instead sweep the adjoint phase $\zeta$ (giving $\bm{x}^{(\ell)} - i (\bm{x}^{(\ell)}_{\mathrm{aj}})^* e^{i\zeta}$) from $0$ to $2 \pi$ repeatedly (e.g., using a sawtooth signal). During the sweep, we record $d_\eta(\zeta)$, the AC component of the measured power monitored through phase shifter $\theta$, $p_{\eta, \mathrm{sum}}(\zeta)$. It is straightforward to show that gradient is $d_\eta(0)$, the AC component evaluated when no adjoint phase is applied (Appendix). To achieve the $\zeta$ sweep physically, we can employ the summing architecture in Fig. \ref{fig:backpropgradient}(c) which sums $\bm{x}^{(\ell)}, i (\bm{x}^{(\ell)}_{\mathrm{aj}})^*$ interferometrically with a constant loss factor of $1 / 2$ in power ($1 / \sqrt{2}$ in amplitude). Then, using a boxcar gated integrator and high pass filter, we can physically compute $d_\eta(\zeta)$ and update the phase shift voltage entirely in the analog domain (Appendix). Ultimately, this approach potentially avoids a costly analog-digital conversion and additional memory complexity required to program $N^2$ elements. 
Since the PNN has $L$ layers, the gradient calculation step requires local feedback circuits at each phase shifter $\eta$ that update the parameters using the measured gradient:
\begin{equation} \label{eqn:gradient}
    \begin{aligned}
    \frac{\partial \mathcal{L}}{\partial \eta} &= -\mathcal{I}(x_\eta x_{\eta, \mathrm{aj}}) \\&= (|x_\eta - i x^*_{\eta, \mathrm{aj}}|^2 - |x_\eta|^2 - |x_{\eta, \mathrm{aj}}|^2) / 2
    \\&= (p_{\eta, \mathrm{sum}} - p_\eta - p_{\eta, \mathrm{aj}}) / 2 = d_\eta(0) / 2,
    \end{aligned}
\end{equation}
where the last equation indicates the equivalence of ``digital subtraction,'' shown in Fig. \ref{fig:backprop} and our proposed ``analog update'' scheme $d_\eta(0) / 2$ in Fig. \ref{fig:backpropgradient}(c, d) (Appendix). Pseudocode and the complete enumerated backpropagation protocol are discussed in the Appendix. Note that the digital and analog gradient updates can both be implemented in parallel across all photonic layers of the network.

\begin{figure*}
    \centering
    \includegraphics[width=\textwidth]{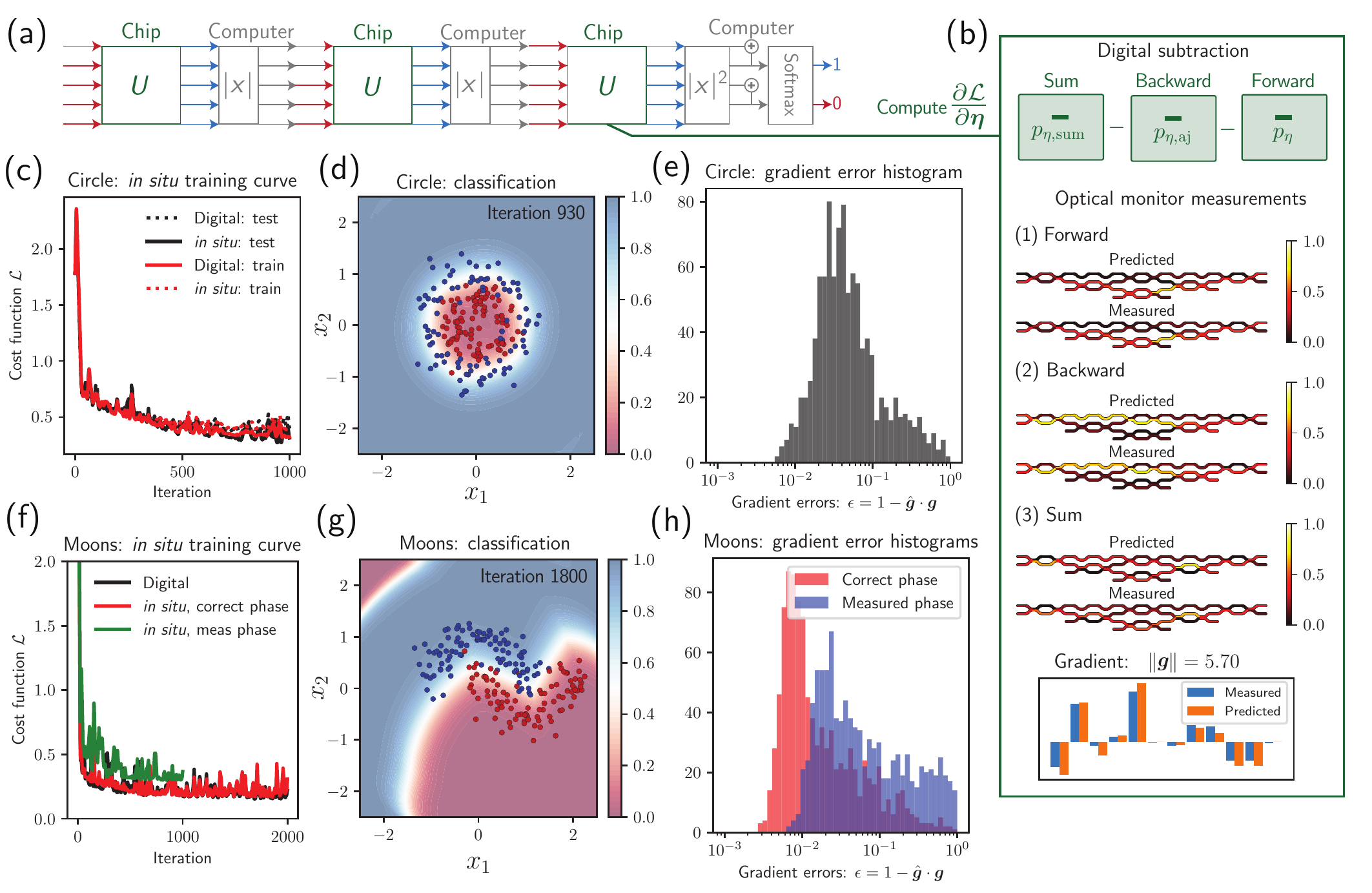}
    \caption{We perform \textit{in situ} backpropagation training based on stochastic gradient descent and Adam update \cite{Kingma2015Adam:Optimization} (learning rate 0.01) for two classification tasks solvable by (a) a three layer hybrid photonic neural network consisting of absolute value nonlinearities and a softmax (effectively sigmoid) decision layer. For the circle dataset, (b) the training curve for Adam update for stochastic gradient descent shows excellent agreement between test and train for both digital and \textit{in situ} backpropagation updates resulting in (c) a classification plot showing the true labels and the classification curve based on the learned parameters resulting in $96\%$ model test accuracy and $93\%$ model train accuracy. (d) For the moons dataset, we compare the test cost curves for the digital and the \textit{in situ} updates, where we find that our phase measurements are sufficiently inaccurate to impact training leading to a lower model train accuracy of $87\%$. If we use ground truth phase measurements (red) instead of measured phases (blue), we ultimately arrive at (e) a sufficiently high model test accuracy of $98\%$ (model train accuracy is lower at $95\%$). When using ground truth phases instead of measured phases, (f) the gradient error reduces considerably by roughly an order of magnitude. (g) While training the circle classification was successful, the measured gradient error is similarly large as that measured in the moons training experiment. This suggests that the importance of accurate gradients can be problem-dependent.}
    \label{fig:backpropexp}
\end{figure*}

Now that we have defined the analog \textit{in situ} backpropagation update, we experimentally evaluate the accuracy of the analog gradient measurement for a matrix optimization problem in Fig. \ref{fig:backpropgradient}(b, d). Since our circuit does not have an explicit backprop unit architecture, we experimentally simulate the ``backprop unit'' of Fig. \ref{fig:backpropgradient}(c) by programming a sequence of summing vectors in the Generator unit of our chip and recording $d_\eta(\zeta)$ to compute the gradient with respect to $\eta$. We implement backpropagation in a single photonic mesh layer optimizing a linear cost function $\mathcal{L}_m = 1 - |\widehat{\bm{u}}_m^T \bm{u}^*_m|^2$, where $\bm{u}_m$ is row $m$ of $U$, a target matrix that we choose to be the four-point discrete Fourier transform (DFT), and $\widehat{\bm{u}}_m$ is row $m$ of $\widehat{U}$, the implemented matrix on the device. Each phase shifter in the photonic network implements a phase shift $\theta + \delta \theta$, where $\theta$ is the optimal phase shift for $U$ and $\delta \theta$ is some random phase error with standard deviation $\sigma_{\theta, \phi}$, which can serve as a measure of ``distance to convergence'' during training of the device. For our gradient measurement step, we send in the derivative $\bm{y}_{\mathrm{aj}} = \frac{\partial{\mathcal{L}_m}}{\partial{\bm{y}}} = -2(\widehat{\bm{u}}_m^T \bm{u}_m^*)^* \bm{e}_m $ to achieve an adjoint field $\bm{x}_{\mathrm{aj}}$, where $\bm{e}_m$ is the $m$th standard basis vector (1 at position $m$, 0 everywhere else). We find in Fig. \ref{fig:backpropgradient}(f) that analog gradient measurement is increasingly less accurate when calculated near convergence, likely due to uncorrected photonic circuit error (e.g. due to loss and/or thermal crosstalk) resulting in large gradient measurement errors.

\section*{Photonic neural net training}

To test overall training within our photonic mesh chip, we assess the accuracy of \textit{in situ} backpropagation in Fig. \ref{fig:backpropexp} to train $L$-layer photonic neural networks to solve multiple 2D classification tasks using the digital subtraction protocol in Ref. \citenum{Hughes2018TrainingMeasurement}. The classification problem assigns points in 2D space to a 0 or 1 label (red or blue coloring) based on whether the point is in a region of space, and the neural network implements the nonlinear boundary (for instance circle-, moon- or ring-shaped) separating points of different labels standardized using the Python package Sklearn and specified in our code \cite{Pai2022Phox:Devices}. The points are randomly synthetically generated and are ``noisy,'' meaning some training example points have a small probability of being assigned a label despite being on the wrong side of the ideal boundary.

To solve this task, we use a three layer PNN where each linear layer uses 4 optical ports ($4 \times 4$ MVM), i.e. $L = 3$ with $N = 4$ inputs and outputs. The inference operation of our photonic neural network consists of programming the inputs in the red Generator circuit and measuring outputs on the blue Analyzer circuit, reprogramming the unitary for each Matrix unit layer on the same chip, and square-rooting the output power measurement to achieve absolute value nonlinearities of the form $|\bm{y}|$ (see Appendix for more detailed description). This unitary layer reprogramming is only intended for a proof-of-concept; the ultimate implementation would dedicate a separate optical device to each linear layer.

The neural network inference model outputs probability of 0 or 1 (red or blue assignment) of each point based on the following model:
\begin{equation} \label{eqn:onn3}
    \begin{aligned}
        \hat{\bm{z}}(\bm{x}) &= \mathrm{softmax2}(|U^{(3)}|U^{(2)}|U^{(1)}\bm{x}|||)
    \end{aligned}
\end{equation}
where we define the $\mathrm{softmax2}: \mathbb{C}^4 \to [0, 1]^2$ as two-element vector representing the probability of 0 or 1 label to be $\mathrm{softmax2}(\bm{y}) = (e^{|y_1|^2 + |y_2|^2}, e^{|y_3|^2 + |y_4|^2}) / (e^{|y_1|^2 + |y_2|^2} + e^{|y_3|^2 + |y_4|^2})$. We then apply a softmax cross entropy (SCE) cost function $\mathcal{L}(\bm{x}) = \mathrm{SCE}(\hat{\bm{z}}(\bm{x}), \bm{z}) = z_0 \log \hat{z}_0 + z_1 \log  \hat{z}_1$. The ultimate goal is to apply automatic differentiation and \textit{in situ} analog gradient measurement on our photonic device to optimize $\mathcal{L}$.

Input data to our device is formatted into the form $(x_1, x_2, p, p)$, where $x_1, x_2$ correspond to the location in 2D space and $p$ is some power ensuring that all inputs are normalized to the same power $P$, i.e. $x_1^2 + x_2^2 + 2p^2 = P$; this convention follows the simple example in Ref. \citenum{Hughes2018TrainingMeasurement}. We perform a 80/20\% train-test split (200 train points, 50 test points), holding out test data from training to ensure no ``overfitting'' takes place, though this is unlikely for our simple model. We generally find higher test than train accuracy in our results since there are fewer ``noisy examples'' in our randomly generated test sets. 

Our single photonic chip is used to perform the data input, data output and matrix operations for all three layers of our photonic neural network shown in Fig. \ref{fig:backpropexp}(a). After each pass through the photonic chip, we measure the output power and digitally perform a square-root operation to effectively implement absolute value nonlinearities on the computer. Throughout the process (forward, backward and sum steps), we also perform digital simulations so we can compare the experimental and simulated performance at each step as shown in Fig. \ref{fig:backprop}(b). Minimizing the cost $\mathcal{L}$ ultimately leads to maximum accuracy in classifying points to the appropriate labels. 

When performing training, the most critical information is in the gradient direction, so we compute gradient direction error using $ 1 - \boldsymbol{g} \cdot \hat{\boldsymbol{g}}$ comparing normalized measured and predicted $\boldsymbol{g} = \partial \mathcal{L} / \partial \bm{\eta} \cdot \|\partial \mathcal{L} / \partial \bm{\eta}\|^{-1}$. An important distinction to make for metric reporting is the difference in ``model'' versus ``device'' cost function and accuracy. In Fig. \ref{fig:backpropexp}, we report ``model metrics'' by evaluating device parameters learned on our chip on the true model. Thus, actual training of the physical parameters is performed on the device itself, and the result of training is evaluated on the computer.

Our first task is to verify that inference works on our platform, which we show for a randomly generated ``ring'' dataset to have 90\% device test set accuracy on our physical platform shown previously in Fig. \ref{fig:backprop}(c). Once we confirm that the inference performance is acceptable, we then perform training of 2D classification problems using our digital subtraction approach on our randomly generated datasets. We use standard gradient update Adam stochastic gradient descent \cite{Kingma2015Adam:Optimization} with a learning rate of 0.01, with all non-linear automatic differentiation performed off the chip via Python libraries JAX and Haiku \cite{Bradbury2022JAX:Programs, TomHennigan2020Haiku:JAX}.

We first report our training update model metrics for the circle dataset for the photonic neural network shown in Fig. \ref{fig:backpropexp}(a). In Fig. \ref{fig:backpropexp}(b), we show the grating tap-to-camera measurements of normalized field magnitudes in the final layer for all three passes required for digital subtraction across all layers of our device at iteration 930 (near the optimum), which show excellent agreement between predicted and measured fields. The training curves in Fig. \ref{fig:backpropexp}(c) indicate that stochastic gradient descent is a highly noisy training process due to the noisy synthetic dataset about the boundary; this phenomenon can be observed for both the digital and analog approaches. Due to these outliers, we only observe convergence in the time (iteration)-averaged curves because even at convergence, updates based on outliers or some incorrectly labelled points can result in large swings in the cost function. These large swings appear roughly correlated between the simulated and measured training curves. Despite these swings and phase shift gradient errors shown in Fig. \ref{fig:backpropgradient}(e), our results of $96\%$ model test accuracy and $93\%$ model train accuracy indicate successful training as shown in Fig. \ref{fig:backpropgradient}(d).

We then train a moons dataset, where we apply the same procedure to achieve a model train accuracy of $87\%$ and model test accuracy of $94\%$, which suggests that training occurs but there is room for improvement as shown in green in Fig. \ref{fig:backpropexp}(f). Upon further investigation, we find that if we use the ground truth phase for the phase measurement but keep the amplitude measurements, we reduce the phase shift gradient error by roughly an order of magnitude on average as shown in Fig. \ref{fig:backpropexp}(h). This results in the successfully trained classification of Fig. \ref{fig:backpropexp}(g) and the red curve in in Fig. \ref{fig:backpropexp}(f) which shows excellent correlation with the black digital training curve. When using the corrected ground truth phase measurement, we achieve a model train accuracy of $95\%$ and model test accuracy of $97\%$ for the full backpropagation demonstration based on measured phase (stopped early at 1000 iterations), an improvement that underscores the importance of accurate phase measurement for improved training efficiency.

\section*{Discussion and outlook}

In this paper, we have laid the foundation for the analysis of our new \textit{in situ} backpropagation proposal and tolerance to gradient errors for the design of practically useful photonic mesh accelerators. Our proof-of-principle experiments suggests that even in the presence of such gradient error, gradient measurement and training photonic neural networks using analog backpropagation updates is efficient and feasible.

Although there exist many approaches for training photonic neural networks, our demonstration and energy calculations (Appendix) suggest that \textit{in situ} backpropagation is the most practical and efficient approach for training deep multilayer hybrid photonic neural networks. Our hybrid approach to training optically accelerates the most computationally intensive operations (both in energy and in time complexity), specifically $O(N^2)$ matrix-vector products and matrix gradient computations (backward pass). On the other hand, all other $O(N)$ computations such as nonlinearities and their derivatives are implemented on the computer directly, which is reasonable because $O(N)$ time is needed to modulate and measure optical inputs and outputs anyway. Other techniques such as population-based methods \cite{Zhang2021EfficientAlgorithm}, direct feedback alignment \cite{Nkland2016DirectNetworks, Filipovich2021MonolithicAlignment}, and perturbative approaches require fewer components to implement but are ultimately less efficient for training deep neural networks compared to backpropagation.

Our main finding is that gradient accuracy plays an important role in reaching optimal results during training. As we find in Fig. \ref{fig:backpropexp}, more accurate gradients result in training convergence speeds and oscillations comparable to digital calculations of the gradients updated over the same training example sequence. This accuracy is vital for \textit{in situ} backpropagation to be a viable competitor to existing purely digital training schemes; in particular, even if individual updates are faster to compute, high error would result in longer training times that mitigate that benefit. In the Appendix, we frame this error scaling in terms of a larger scale PNN simulation on the MNIST dataset originally as explored in Ref. \cite{Williamson2020ReprogrammableNetworks}, where we consider errors in gradient measurement due to optical I/O errors and photodetector noise at the global monitoring taps.

Our findings ultimately have wide ranging implications because backpropagation is the most efficient and widely used neural network training algorithm in conventional machine learning hardware used today. Our analog approach for machine learning thus opens up a vast opportunity for energy-efficient artificial intelligence applications using photonic hardware. We additionally provide seamless integration into current machine learning training protocols (e.g. autodifferentiation frameworks such as JAX \cite{Bradbury2022JAX:Programs} and TensorFlow \cite{Abadi2016TensorFlow:Learning}). A particularly impactful opportunity is in data center machine learning where optical signals already store data that can be fed into PNNs for inference and training tasks. In such settings, our demonstration presents a key new opportunity for both inference and training of hybrid PNNs to dramatically reduce carbon footprint and counter the exponentially increasing costs of AI computation.

\section*{Acknowledgements}
We would like to acknowledge Advanced MicroFoundries (AMF) in Singapore for their help in fabricating and characterizing the photonic circuit for our demonstration and Silitronics for their help in packaging our chip for our demonstration. We would also like to acknowledge funding from Air Force Office of Scientific Research (AFOSR) grants FA9550-17-1-0002 in collaboration with UT Austin and FA9550-18-1-0186 through which we share a close collaboration with UC Davis under Dr. Ben Yoo. Thanks also to Payton Broaddus for helping with wafer dicing, Simon Lorenzo for help in fiber splicing the fiber switch for bidirectional operation, Nagaraja Pai for advice on electrical and thermal control packaging, and finally Carsten Langrock and Karel Urbanek for their help in building our movable optical breadboard.

\section*{Data and software}
All software and data for running the simulations and experiments are available through Zenodo \cite{Pai2022Solgaardlab/photonicbackprop:Networks} and Github through the Phox framework, including our experimental code via Phox \cite{Pai2022Phox:Devices}, simulation code via Simphox \cite{Pai2022Simphox:Library}, and circuit design code via Dphox \cite{Pai2022Dphox:Design}.

\section*{Contributions}
SP taped out the photonic integrated circuit and ran all experiments with input from ZS, TH, TP, BB, NA, MM, OS, SF, DM. SP and ZS wrote code to control experimental device. TP designed the custom PCB with input from SP. SP wrote the manuscript with input from all coauthors. All coauthors contributed to discussions of the protocol and results.

\section*{Conflicts of interest}
SP, ZS, TH, IW, MM, SF, OS, DM have filed a patent for the analog backpropagation update protocol discussed in this work with Prov. Appl. No.: 63/323743. The authors declare no other conflicts of interest.

\section*{Methods}

\subsection{Circuit design and packaging}
Our photonic integrated circuit is a $6 \times 6$ triangular photonic mesh consisting of a total of $15$ MZIs fabricated at the AdvancedMicroFoundry (AMF) in Singapore designed using our photonic library DPhox \cite{Pai2022Dphox:Design} which is a custom automated photonic design library in Python. Each of the MZIs in the mesh is controlled using programmable phase shifters in the form of $80\ \mu$m $\times$ $2\ \mu$m titanium nitride heaters with $10.5$ ohm/sq sheet resistance surrounded by deep trenches that are $80\ \mu$m $\times$ $10\ \mu$m and a total of $7 \mu$m away from the waveguide, which use resistive heating to control the interference of light propagating in the chip. The MZIs consist of two 50/50 directional couplers, with S-bends consisting of $30$ $\mu$m radius arc turns and $40\ \mu$m long interaction lengths with a $300$ nm gap. Next to each of the phase shifters is a bidirectional grating tap monitor, which is a directional coupler tap that couples 3\% of the light propagating either forward or backward through the waveguide attached to the tap and feeds that light to a grating to be imaged on a camera focused on the grating. Traces for one of the terminals of each of the phase shifters are routed to separate individual pads on the edge of the chip, and the ground connections across all phase shifters in a column of MZIs are shared and connected to a single ground pad. The trace widths need to be thick enough to handle high thermal currents, so we use 15 $\mu$m wide traces and $15 N_{\mathrm{wire}}$ $\mu$m wide traces when multiple connections are connected to a shared ground contact.

The photonic chip is attached using silver paint to a 1.5mm thick copper shim and a custom Advanced Circuits PCB designed in KiCAD consisting of ENIG coated metal traces to interface the phase shifters with an NI PCIe-6739 controller for setting programmable phase shifts throughout the device. Our PCB is wirebonded using two-tier wirebonding to the chip by Silitronics Solutions, made possible by fanout to NI SCB-68 connectors that interface directly to our PCIe-6739 system. The input optical source is a Agilent 81606A tunable laser with a tunable range of 1460 nm to 1580 nm. The laser light is coupled into a single-mode fiber and optically interfaced to the chip using W2 Optronics 127 micron pitch fiber array interposers at the left and right sides of the mesh, with a mirror facet designed to couple optical signals at 10 degrees from the normal as we only need to couple into a single grating coupler for each fiber array coupler. Optical stray reflections from light not coupled into the chip generally interfere with grating tap signals forming extra streaks in the camera; these stray reflections are blocked using pieces of paper carefully placed above the fiber arrays that act as lightweight removable stray light blockers. 

For thermal stability, this chip-PCB assembly is thermally connected to a thermoelectric cooler (TEC). This thermal connection is made possible by metal vias connecting rectangular ENIG-coated copper patches on the top of the PCB to the bottom of the PCB, with thermal paste between an aluminum heat sink mount and the bottom rectangular metal patch. For feedback control, a thermistor placed near the chip and the TEC under the chip are attached to a TEC controller unit, allowing stable chip temperature (kept at 30$^\circ$C) for training.

\begin{figure*}
    \centering
    \includegraphics[width=\textwidth]{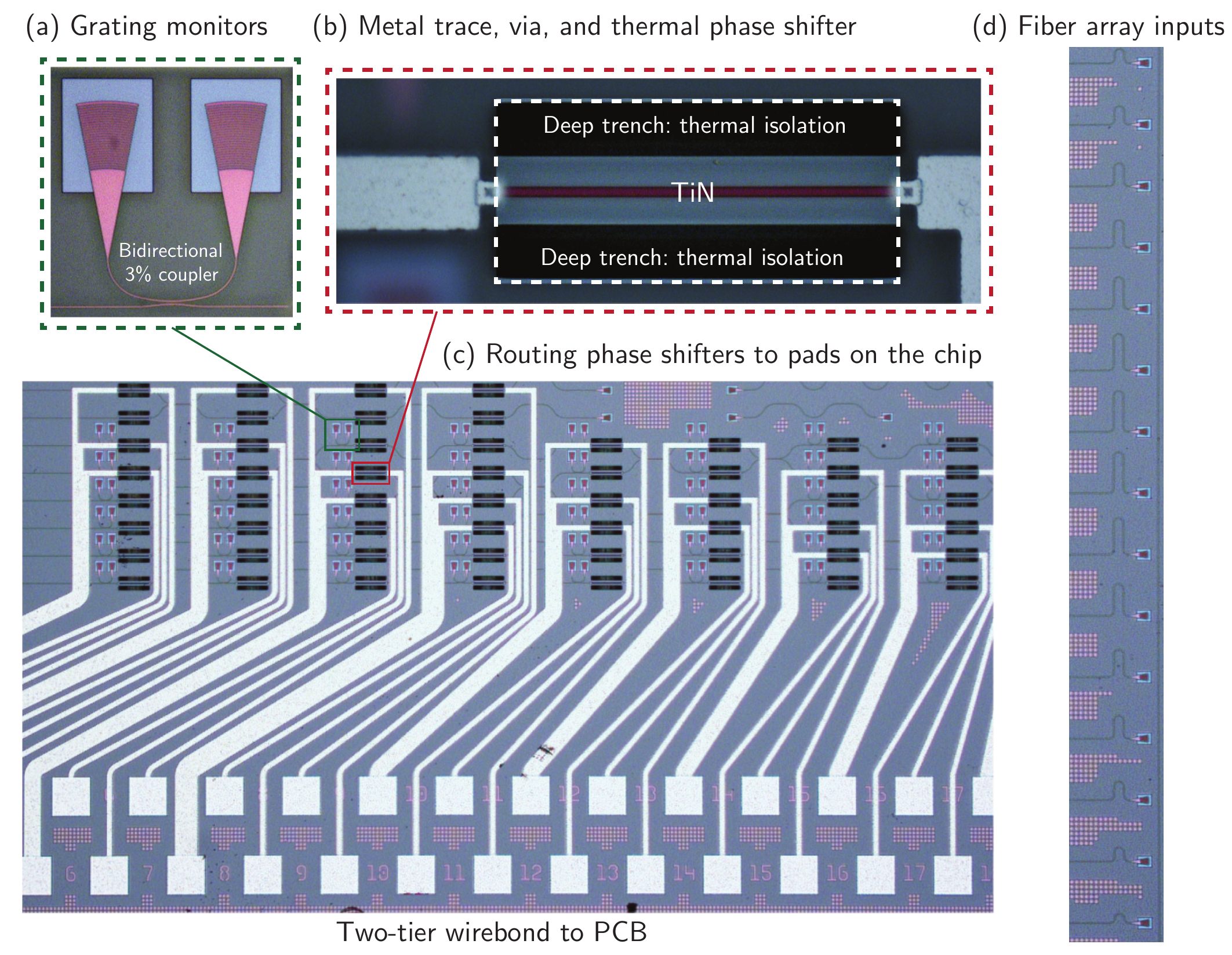}
    \caption{Microscope images of the photonic mesh used in this paper. (a) Grating monitor closeup showing the bidirectional grating tap we use to perform the backpropagation protocol. (b) Metal trace, via, and TiN (titanium nitride) phase shifter is colocated with the grating monitor and is used to control the interference by changing optical phase in the mesh programmatically. Deep trenches are used for thermal isolation. Here, we show an overlay of phase shifter focal plane on the top metal trace and via used to connect each phase shifter to the pads. (c) A large scale view of a section of the chip (d) Fiber array inputs to the photonic mesh are spaced 127 $\mu$m apart and are used for interfacing fiber arrays.}
    \label{fig:meshimages}
\end{figure*}

\subsection{Optical rig design}
Our optical rig consists of an Ethernet cable-connected Xenics Bobcat 640 IR camera and microscope assembly mounted on an XY stage and six-axis stages for free space fiber alignment. The IR camera and microscope image individual grating taps throughout a photonic integrated circuit (PIC) and is responsible for all measurement on the chip (both optical I/O and optical gradient monitoring).

The microscope uses an $\infty$-corrected Mitutuyo IR 10x objective and a 40cm tube lens leading to a dichroic connected to visible and IR optical paths for simultaneous visible and infrared imaging. The optical rig is also outfitted with additional paths for LEDs to illuminate the actual chip features. This allows us to find the optimal focus for the grating spots, an image shown in Fig. \ref{fig:backpropgradient}(a). In order to measure intensities directly using the IR camera, the Bobcat camera ``Raw'' mode is turned on and autogain features are turned off. The integration time is set to 1 millisecond, and the input laser power is set to 3 mW; note that higher integration times are required for lower input laser powers. We take an initial reference image to get a baseline and then to measure the spots intensities or powers, we sum up the pixel values that ``fill'' the appropriate grating taps throughout the device. The triangular mesh circuit is constructed such that the grating taps lie along columns of devices, which means the optical rig images a $6 \times 19$ array of spots. The infrared path has roughly a $700 \times 600$ $\mu$m field of view, allowing simultaneous measurements of $6 \times 3$ grating spots on the chip (MZIs are $625\ \mu$m long in total given roughly $165\ \mu$m long directional couplers), which necessitates an XY translation stage to image multiple spots simultaneously on the chip.

The speed of backpropagation is limited by the mechanics of the XY stage required to image spots throughout the chip, so our demonstration training experiments took up to 31 hours of real time to run, limited primarily by the wait time for the stage to settle on various groups of spots on the chip. Assuming $T$ iterations, the stage needs to move a total of $15T$ times (5 for each of the three \textit{in situ} backpropagation steps to be able to image all of the spots). For 1000 iterations, the stage needs to move a total of $15000$ times which necessitates the need of automation for the stage of our proof-of-concept demonstration. In a final commercial implementation, the grating taps would be replaced by integrated photodetectors; there would in principle be no separate optical rig system in a fully packaged hybrid digital-analog photonic circuit.

\subsection{Forward inference operation}

Forward inference proceeds as follows for layer $\ell$ (see Fig. \ref{fig:backprop}(a) in the main text) where each step is $O(N)$:
\begin{enumerate}
    \item Compute the sets of phase shifter settings $\bm{\theta}^{(\ell)}_X, \bm{\phi}^{(\ell)}_X$ for the Generator to give the desired vector $\bm{x}^{(\ell)}$ of complex input amplitudes for the Matrix unit in layer $\ell$ .
    \item Set these as the actual phase shifts in the Generator phase shifters using calibration curves for $v_\theta(\theta), v_\phi(\phi)$ and shine light into the Generator circuit to create the corresponding actual vector of optical input amplitudes for the Matrix unit.
    \item After the propagation of light through the Matrix unit, the system has optically evaluated the vector of complex optical output amplitudes $\bm{y}^{(\ell)} = U^{(\ell)} \bm{x}^{(\ell)}$. Now self-configure \cite{Miller2013Self-aligningCoupler} the output Analyzer circuit to give all the output power in the ``top'' output waveguide, and note the corresponding sets of voltages $\bm{v}_\phi$ and voltages $\bm{v}_\theta$ now applied to each phase shifter in the Generator circuit.
    \item Deduce the phase shifts $\bm{\theta}^{(\ell)}_Y, \bm{\phi}^{(\ell)}_Y$ in the Analyzer circuit using calibration curves for $\theta(v_\theta), \phi(v_\phi)$, and hence compute the corresponding measured output amplitudes $\bm{y}^{(\ell)}$ .
    \item Compute $\bm{x}^{(\ell + 1)} = f^{(\ell)}(\bm{y}^{(\ell)})$ on the computer.
\end{enumerate}
The first four steps are also used in cases where light is sent backwards (see Fig. \ref{fig:backprop}(g, h)), switching the role of the input and output vector units from Generator to Analyzer and vice versa. Pseudocode for the forward operation of the PNN is provided in the Appendix, and code for the actual implementation is provided in our photonic simulation and control framework Phox \cite{Solgaardlab/phox:Devices}.

\subsection{Backpropagation protocol}

For each training example $(\bm{x}, \bm{z})$, we calculate gradient updates to phase shifts $\bm{\eta}$ using a ``backward pass'' corresponding to the inference ``forward pass'' for that data. More formally, we define a ``vector-Jacobian product'' or VJP for each function $U^{(\ell)}, f^{(\ell)}$ to algorithmically compute the gradient of our cost function $\mathcal{L}$. As shown in Fig. \ref{fig:backprop}(c), each transformation from the forward step is mapped to a VJP in the corresponding backward step (defined in decreasing order from layer $L$ to $1$) which depends on intermediate function evaluations in both forward and backward passes. The \textit{in situ} backpropagation step implements the costly intermediate VJP evaluations (i.e. matrix multiplications) directly in the analog optical domain. We define the VJP for nonlinearity $f^{(\ell)}(\bm{y}^{(\ell)})$ as $f_{\mathrm{vjp}}^{(\ell)}(\bm{y}^{(\ell)}, \bm{x}_{\mathrm{aj}}^{(\ell + 1)})$:
\begin{equation} \label{eqn:vjp}
    \begin{aligned}
        \bm{y}_{\mathrm{aj}}^{(\ell)} &= f_{\mathrm{vjp}}^{(\ell)}(\bm{y}^{(\ell)}, \bm{x}_{\mathrm{aj}}^{(\ell + 1)}) \\
        \bm{x}_{\mathrm{aj}}^{(\ell)} &= (U^{(\ell)})^T \bm{y}_{\mathrm{aj}}^{(\ell)}
    \end{aligned}
\end{equation}

Finally, we synthesize Eqs. \ref{eqn:onn} and \ref{eqn:vjp} and the results of Ref. \citenum{Hughes2018TrainingMeasurement} to get the backpropagation update based on applying the chain rule evaluating the cost function at a random training example $\bm{x}_t, \bm{z}_t$ at iteration $t$:
\begin{equation}
    \begin{aligned}
        \frac{\partial \mathcal{L}}{\partial \bm{\eta}^{(\ell)}} &= \frac{\partial \bm{y}^{(\ell)}}{\partial \bm{\eta}^{(\ell)}} \overbrace{\frac{\partial \bm{x}^{(\ell + 1)}}{\partial\bm{y}^{(\ell)}} \cdots \underbrace{\frac{\partial \widehat{\bm{z}}}{\partial\bm{y}^{(L)}} \frac{\partial \mathcal{L}}{\partial \widehat{\bm{z}}}}_{\bm{y}^{(L)}_{\mathrm{aj}}}}^{\bm{x}^{(\ell)}_{\mathrm{aj}}} \Bigg|_{\bm{x}_t, \bm{z}_t}
        \\ &= \overbrace{\frac{\partial \bm{y}^{(\ell)}}{\partial \bm{\eta}^{(\ell)}}}^{D_\ell \times N \mathrm{\ Jacobian}} \cdot \overbrace{\bm{x}^{(\ell)}_{\mathrm{aj}}}^{N \times 1 \mathrm{\ vector}} \\&= \overbrace{(\bm{x}^{(\ell)})^T \frac{\partial U^{(\ell)}}{\partial \bm{\eta}^{(\ell)}}\bm{x}^{(\ell)}_{\mathrm{aj}}}^{``\mathrm{optical \ VJP}"} =  \overbrace{-\mathcal{I}(x_{\bm{\eta}^{(\ell)}} x_{\bm{\eta}^{(\ell)}, \mathrm{aj}})}^{D_\ell \times 1 \mathit{\ in\ situ}\ \mathrm{gradient}} \\
        \bm{\eta}_t &:= \bm{\eta}_{t - 1} + \alpha \frac{\partial \mathcal{L}}{ \partial \bm{\eta}} \Bigg|_{\bm{x}_t, \bm{z}_t}
    \end{aligned}
\end{equation}
where $x_{\bm{\eta}}$ represents a vector of intermediate fields at the input of phase shifters in layer $\ell$ $\bm{\eta}^{(\ell)}$ at iteration $t$, $D_\ell$ is the number of phase shifts parametrizing the device at layer $\ell$, $\mathcal{I}$ refers to imaginary part, and $\alpha$ is the learning rate. The main idea is that if enough training examples are supplied (i.e., after $T$ updates), the device will automatically discover or ``learn'' a function that performs the task we desire. 

Based on Eq. \ref{eqn:vjp}, the steps of our optical VJP step, as depicted in Fig. \ref{fig:backprop}(c), is as follows in order from layer $\ell = L$ to $1$ of the photonic neural network:
\begin{enumerate}
    \item Compute the ``adjoint'' vector $\bm{y}_{\mathrm{aj}}^{(\ell)} = f_{\mathrm{vjp}}^{(\ell)}(\bm{y}^{(\ell)}, \bm{x}_{\mathrm{aj}}^{(\ell + 1)})$. For the last layer, set $\bm{y}_{\mathrm{aj}}^{(\ell)}$ to be the \textit{error signal} $\bm{y}_{\mathrm{aj}}^{(L)} = \left(\frac{\partial \mathcal{L}}{\partial \bm{x}^{(L + 1)}}\right)^*$.
    \item Perform the backward ``adjoint'' pass $\bm{x}_{\mathrm{aj}}^{(\ell)} = U^T \bm{y}_{\mathrm{aj}}^{(\ell)}$ by sending light backwards through layer $\ell$ of the mesh and measuring the resulting vector of amplitudes $\bm{x}_{\mathrm{aj}}^{(\ell)}$ emerging backwards from the mesh.
    \item Send the vector of optical amplitudes $\bm{x}^{(\ell)} - i (\bm{x}^{(\ell)}_{\mathrm{aj}})^*$ forward into layer $\ell$ of the mesh. 
    \item Measure gradient $\partial \mathcal{L} / \partial \theta$ for any phase shifter $\theta$:
    \begin{enumerate}
        \item If using digital subtraction measurement \cite{Hughes2018TrainingMeasurement}, measure the sum power $p_{\theta, \mathrm{sum}}$ and subtract $p_\theta$ and $p_{\theta, \mathrm{aj}}$ (monitored power from forward and backward steps) to get the gradient. 
        \item If using analog gradient measurement, sweep the adjoint global phase $\zeta$ (giving $\bm{x}^{(\ell)} - i (\bm{x}^{(\ell)}_{\mathrm{aj}})^* e^{i\zeta}$) from $0$ to $2 \pi$ repeatedly (e.g., using a sawtooth signal). Measure $d_\theta(\zeta)$, the AC component of the measured power through phase shifter $\theta$, $p_{\theta, \mathrm{sum}}(\zeta)$. The gradient is $d_\theta(0) / 2$.
    \end{enumerate}
    \item Update $\bm{\eta}$ using measured gradients $\partial \mathcal{L} / \partial \bm{\eta}$.
\end{enumerate}
Note that Step 1 can be simplified to $\bm{y}_{\mathrm{aj}}^{(\ell)} = (f^{(\ell)})'(\bm{y}^{(\ell)}) \odot \bm{x}_{\mathrm{aj}}^{(\ell + 1)}$ in the case that $f^{(\ell)}$ is holomorphic, or complex-differentiable. In this paper for the neural network parametrized by Eq. \ref{eqn:onn3}, we specifically care about the nonlinearity $f^{(\ell)}(\bm{y}) = |\bm{y}|$, which has the associated VJP:
\begin{equation}
    f_{\mathrm{vjp}}^{(\ell)}(\bm{y}, \bm{x}_{\mathrm{aj}}) = \frac{\bm{y}}{|\bm{y}|} \cdot \mathcal{R}(\bm{x}_{\mathrm{aj}}).
\end{equation}
The other VJP required to calculate $\partial \mathcal{L} / \partial \bm{y}^{(L)}$ from the final softmax cross entropy and power measurement at the end of the network is handled by our automatic differentiation framework JAX \cite{Bradbury2022JAX:Programs, TomHennigan2020Haiku:JAX}.

Steps 3 and 4 can be parallelized over all layers (i.e., parameters of the network) for both the digital and analog update schemes. Pseudocode for the overall protocol (using digital subtraction), along with an energy-efficient proposal for analog gradient computation, is discussed in the Appendix. The final step can be achieved using ``stochastic gradient descent'' (which independently updates the loss function based on randomly chosen training examples) or adaptive learning where the update vector depends both on past updates and the new gradient. A successful and commonly used implementation of this, which we use in this paper, is called the Adam update \cite{Kingma2015Adam:Optimization}.

\appendix

\section{Energy and latency analysis}

In this section, we justify why the analog \textit{in situ} update discussed in the main text may be chosen over the digital update proposed in Ref. \cite{Hughes2018TrainingMeasurement} and used in our main backpropagation training demonstration.

In our hybrid scheme, most of the computation is concentrated in sending in $N$ input modes using modulators (each taking energy $E_{\mathrm{inp}}$) and digital-analog converters and measuring the $N$ output mode powers and amplitudes using photodetectors and analog-digital converters (each taking energy $E_{\mathrm{meas}}$). Therefore, the various approaches for a given matrix-vector product cost roughly $N \cdot (E_{\mathrm{inp}} + E_{\mathrm{meas}})$, equivalent to the cost for setting up the input/output behavior for the photonic mesh. A digital electronic computer, on the other hand, requires $N^2$ sequential operations (i.e., multiply-and acumulate operations that are \textit{not} parallel) to compute any given matrix-vector product. 

Beyond inference tasks, the additional backward and sum steps required for \textit{in situ} backpropagation adds additional energy and latency contributions. The analog update explored in Fig. \ref{fig:backpropgradient} requires $N^2$ optoelectronic units for energy-efficient operation, each of which is outfitted with a photodetector, a lock-in amplifier, and high-pass filter consuming energy $E_{\mathrm{grad}}$ to measure $d_\theta(0)$ for a total energy consumption of $N^2 E_{\mathrm{grad}} + N \cdot (3E_{\mathrm{inp}} + 2E_{\mathrm{meas}})$ for all three steps of the full backpropagation measurement. The $2E_{\mathrm{meas}}$ comes from the output measurements in the first two steps, and the $E_{\mathrm{grad}}$ comes from an analog gradient measurement in the final step.

In comparison, the digital subtraction described in Fig. \ref{fig:backprop} can be useful in adaptive updates that require storing information about previous gradients (such as Adam \cite{Kingma2015Adam:Optimization} which we exploit for training), but there are a couple of drawbacks. First, a digital update is less memory efficient since $N^2$ elements need to be stored using analog memory to be able to run the ``digital subtraction'' computation in backpropagation. Additionally, the total energy consumption becomes $3N^2 E_{\mathrm{grad, digital}} + N \cdot (3E_{\mathrm{inp}} + 2E_{\mathrm{meas}})$, with $E_{\mathrm{grad, digital}} \gg E_{\mathrm{grad, analog}}$ due to large numbers of analog-digital conversions required to implement the analog-digital conversions and digital subtraction calculations. Analog-digital conversions are among the most energy- and time-consuming operations in a hybrid photonic device; when operating at GHz speeds, the best individual comparators generally require up to $40$ fJ \cite{Filippini2018AConversion, Miyahara2008AADCs} (versus around $1$ fJ/bit for input modulators \cite{Wang2018NanophotonicModulatorsb}) and therefore should ideally be reserved for optical input/output in the photonic meshes.

A final energy consideration is the phase shift modulation. These voltage-controlled modulators may be controlled by thermal actuation \cite{Harris2014EfficientSilicon}, microelectromechanical (MEMS) actuation \cite{Errando-Herranz2020MEMSCircuits} or phase-change materials such as barium titanate (BTO) \cite{Wuttig2017Phase-changeApplications}. Of these options, MEMS actuation is among the most promising because unlike thermally actuated phase shifters, they cost no energy to maintain a given programmed state (``static energy''), dramatically improving the energy efficiency of operation compared to thermal phase shifters which constantly dissipate large amounts of heat. Additionally, unlike phase change materials, MEMS phase shifters use CMOS materials such as silicon or silicon nitride. Furthermore, such devices can be designed to operate in the linearly with voltage \cite{Edinger2019Low-lossPhotonics, Edinger2020CompactPhotonics} which ensures that the gradient update applied to the voltage is the same as that of the phase shift without a calibration curve. This helps with gradient accuracy as we discuss now.

\section{Gradient accuracy}

\begin{figure*}
    \centering
    \includegraphics[width=\textwidth]{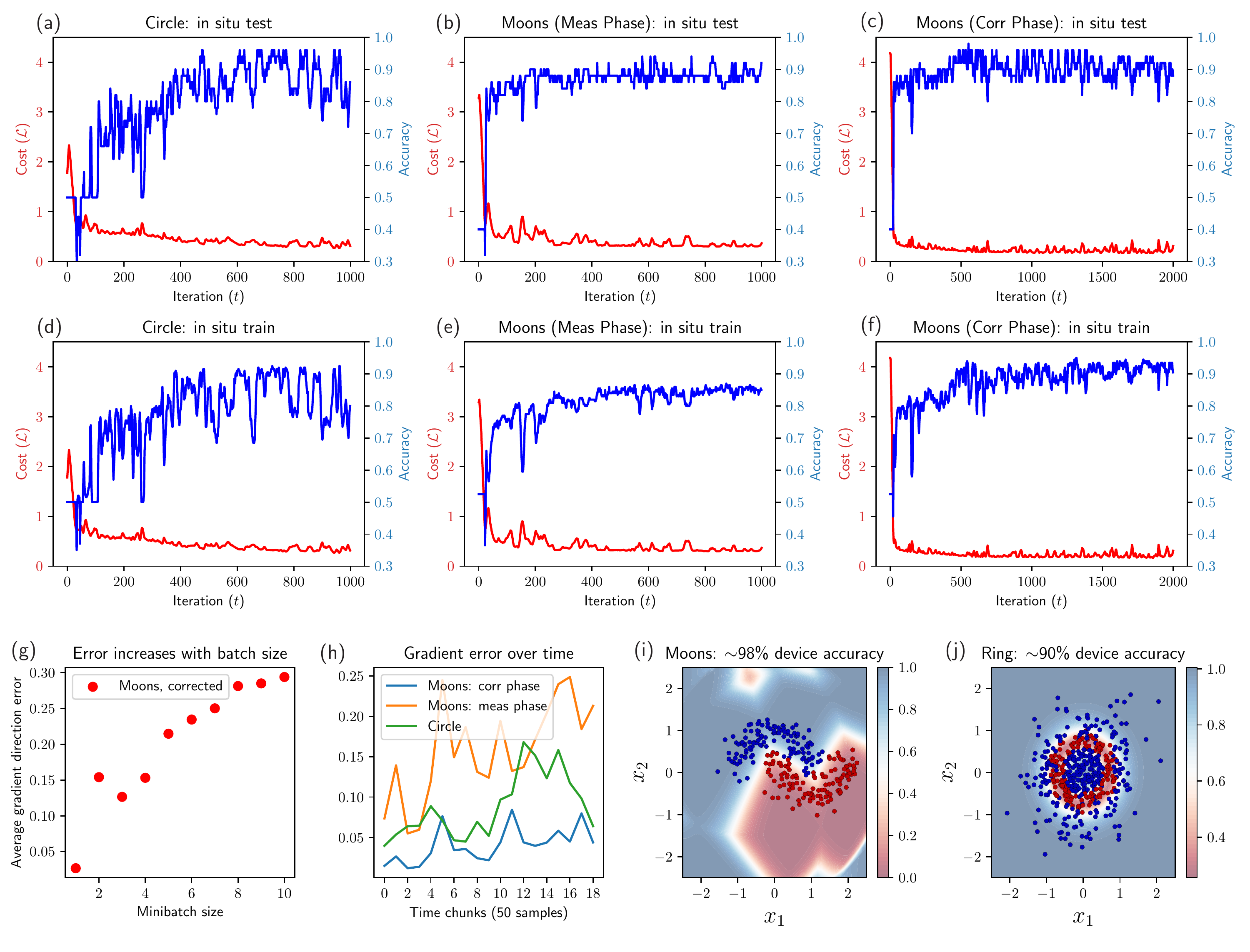}
    \caption{(a-f) A comparison of the model cost and accuracy curves between circle, moons (measured) and moons (corrected) experiments comparing test (a-c) and train (d-f) data. (g) The error in the gradient increases with the batch size. (h) The gradient error increases over the course of the optimization here shown as a time-averaged series averaged over 50-sample time (iteration) chunks. This increase has to do with the fact that near convergence, the measured gradient is smaller, leading to slightly larger errors. (i, j) Device accuracy for moons and ring dataset inference tasks showing the model boundary (calculated on the computer) in the background and device-classified points (in red, blue). With this evidence, we make an assumption that model and device metrics are close enough, so we use model metrics throughout the paper.}
    \label{fig:accuracy}
\end{figure*}

As shown in Fig. \ref{fig:backpropexp}(f, h) and in Fig. \ref{fig:accuracy}(h), gradient accuracy can affect the optimization and decrease as the optimization approaches convergence. As we find in the main text, accurate phase measurement plays an important role in measuring accurate gradients. This is true even when the nonlinearity (as in our case with absolute value) removes the need to measure phases in the inference step. Since in the main text, we evaluate the model accuracy (device-trained parameters evaluated on a theoretical computer model), we also show some evidence that the device and model classifications match quite well in Fig. \ref{fig:accuracy}(i, j).

\begin{figure*}
    \centering
    \includegraphics[width=\textwidth]{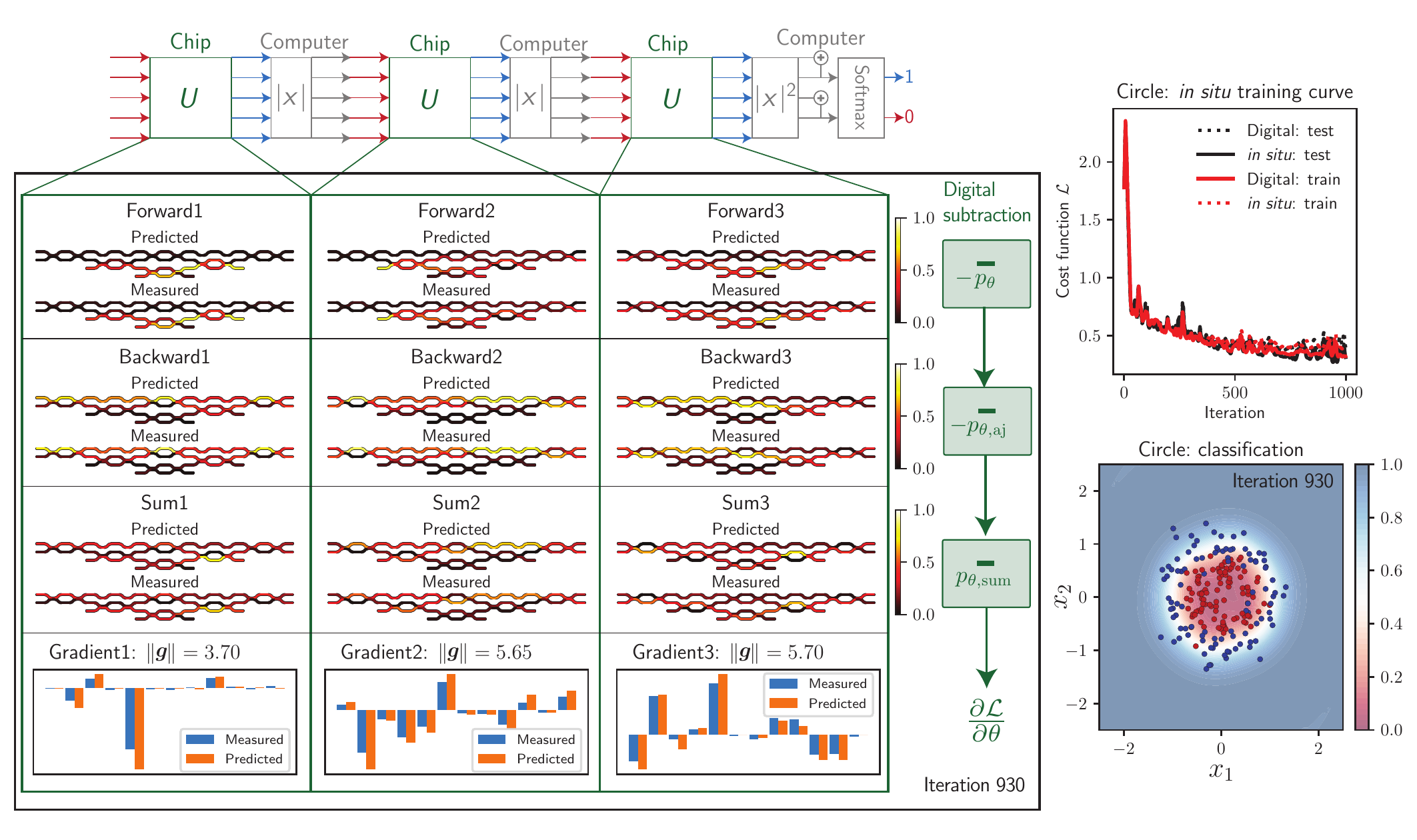}
    \caption{Here, we simply extend the data shown for iteration 930 in Fig. \ref{fig:backpropexp}, specifically showing the intermediate power measurements at each point in the photonic neural network. As indicated by ``digital subtraction,'' we directly subtract the sum measurements by the top forward and backward measurements.}
    \label{fig:nniter}
\end{figure*}

One popular type of update is based on ``minibatch gradient descent,'' a machine learning technique that calculating gradients based on multiple training examples. This would dramatically smooth out the noisy training curves shown in Fig. \ref{fig:accuracy}(a-f), as the resulting averaged gradients would actually be much smaller in magnitude. However, as shown in Fig. \ref{fig:accuracy}(g), we find that the normalized error of a minibatch gradient is generally significantly higher than that of the gradient for a single training example which can have negative implications for training. This is because the variance of the gradient error remains the same when averaged over many examples, but the contribution of the gradient error is much larger over a batch. This phenomenon might be problem-dependent; if the average gradient for the minibatch is not closer to zero than the gradient for individual training examples, this error may not be an issue. This underscores the importance of accurate gradient measurement, which can be improved using more accurate output phase measurements; our output phase measurement alone results in an order-of-magnitude increase in gradient error.

Finally, a linear relationship between phase and voltage can help to improve gradient update accuracy without requiring nontrivial scaling complexity in the hardware. In other words, we ensure $\partial \mathcal{L} / \partial v_\theta = \partial \mathcal{\theta} / \partial v_\theta \cdot \partial \mathcal{L} / \partial \theta$ with constant $\partial \mathcal{\theta} / \partial v_\theta$ which simplifies the required analog circuitry. The $\partial \mathcal{\theta} / \partial v_\theta$ term is calculated using calibration curves, and this assumption is more-or-less valid in our case as we operate the phase shifters in the linear regime as shown in Fig. \ref{fig:calibration}(e).

\section{Usage in machine learning software}

Backpropagation is also known as automatic differentiation (AD) because any program that uses backpropagation registers a ``backward'' gradient function for any forward function, which is used by AD Python engines such as JAX\cite{Bradbury2022JAX:Programs}, TensorFlow2 \cite{Abadi2016TensorFlow:Learning}, and PyTorch. 
We demonstrate that our protocol can be easily coupled with an existing automatic differentiation framework (JAX and Haiku \cite{Bradbury2022JAX:Programs, TomHennigan2020Haiku:JAX}), which can register a backward step and adaptive update based on Adam \cite{Kingma2015Adam:Optimization} for all unitary matrix operations as an analog \textit{in situ} backpropagation gradient calculation rather than an expensive digital operation.
In this way, the digital side of our hybrid PNN never needs to store or have any knowledge of parameters in the photonic mesh architecture.
However, in cases where adaptive gradient updates are used, such as Adam, aggregated knowledge based on past gradient updates needs to be stored; non-volatile memory may be required to energy-efficiently store these additional parameters.

\section{Comparison with other training algorithms}
Backpropagation is the most widely used and efficient known algorithm for training multilayer neural network models, though it is far from the only method for calculating gradient-based updates.

Finite differences-based training has been proposed as a method of training photonic neural networks \cite{Shen2017DeepCircuits}. 
Finite differences falls under the umbrella of perturbative learning, a well known and model-free analog machine learning technique for analog neural networks that works by perturbing each element by a small amount, or perturbing many elements simultaneously, and measuring the resulting change in the overall loss function \cite{Dembo1990Model-FreeLearning, Cauwenberghs1992AOptimization, Alspector1992ANetworks}. 
Perturbative learning is most useful in the context of \textit{fully} optical neural networks that implement nonlinearities directly on the device, an example of which has previously been proposed for all-photonic neural networks \cite{Williamson2020ReprogrammableNetworks}. 
The PNN architecture relevant for perturbative learning is therefore fundamentally different from the hybrid PNN we propose that can benefit from \textit{in situ} backpropagation. 

It is worth noting that hybrid PNNs can be more versatile and useful to a larger range of traditional AI applications compared to fully analog PNNs \cite{Williamson2020ReprogrammableNetworks}.
Many complex models (e.g. as transformers, convolutional networks, word embedding layers and recurrent neural networks) used in machine intelligence today are more easily implemented in hybrid rather than all-analog systems due to the sheer complexity and logic implemented in the model architectures.

Additionally, backpropagation is significantly more efficient than finite differences and other similar adaptive approaches.
In backpropagation, the time complexity of the ``forward-propagated'' inference pass or direct evaluation of the model is roughly the same as that of the ``backpropagated'' gradient calculation pass.
In contrast, a perturbative gradient calculation is significantly more costly since it cannot be computed on a layer-by-layer basis; the forward propagation must continue on to the end of the network, which does not favor our hybrid approach.

Other alternatives to backpropagation include direct-feedback alignment (DFA) \cite{Nkland2016DirectNetworks, Filipovich2021MonolithicAlignment}, derivative-free optimization and population-based learning, which include evolutionary-based (genetic algorithm or GA) \cite{Zhang2021EfficientAlgorithm} and swarm-based methods. The GA and DFA training approaches have been recently experimentally demonstrated to successfully train optical devices at moderately challenging machine learning tasks \cite{Filipovich2021MonolithicAlignment, Zhang2021EfficientAlgorithm}. However, these are generally regarded to be less efficient at training models compared to backpropagation and are have not proven to scale to more challenging image and word processing machine learning benchmarks like ImageNet \cite{Deng2010ImageNet:Database}. Work is still required to test the scalability of photonic machine learning to solve problems of such complexity as ImageNet.

\section{Photonic mesh operation}

\subsection{Bidirectional matrix multiplication}

In photonic neural networks, programmable photonic meshes act to perform compute-intensive linear operations that preserve the overall power in the form of unitary transmission operator $U$. Meshes are configured using three subunits: an input vector generator network (generating $\bm{x}$), a matrix network (multiplying by $U$), and an output vector analyzer network (measuring $\bm{y}$). Our mesh is ``bidirectional'' in the sense that it can represent matrix-vector operations regardless of whether the light is shined in the forward (left-to-right) or backward (right-to-left) direction as depicted in Fig. \ref{fig:backprop}(a) of the main text, where in the latter case the output analyzer and input generator switch places.

\subsection{Tunable splitter}
A tunable splitter, the basic building block of a photonic mesh, is a $2 \times 2$ element that consists of a tunable split ratio region and a differential phase shifter at the input or output. For straightforward calibration, we may use Mach-Zehnder interferometer building blocks that consist of a differential $\phi$ phase shift, 50/50 splitter, differential $\theta$ phase shift, and then a final 50/50 splitter, giving us the following mathematical representation acting on modes $x_1, x_2$ and yielding outputs $y_1, y_2$:
\begin{equation} \label{eqn:tb}
    \begin{aligned}
    \begin{bmatrix}y_1 \\ y_2 \end{bmatrix} &= i\begin{bmatrix}e^{i\phi}\sin \frac{\theta}{2} & \cos \frac{\theta}{2} \\ e^{i\phi}\cos \frac{\theta}{2} & -\sin \frac{\theta}{2}
    \end{bmatrix} \begin{bmatrix}x_1 \\ x_2 \end{bmatrix}\\
    \bm{y} &= T_2(\theta, \phi) \bm{x},
    \end{aligned}
\end{equation}
where $\theta \in [0, \pi]$ and $\phi \in [0, 2\pi)$. In practice, due to the nonlinear relationship between the phase shifts $\theta, \phi$ and the respective voltage drives $v_\theta, v_\phi$, we instead may need to represent $T_2$ with an additional global phase as:
\begin{equation} \label{eqn:tbupper}
    \widetilde{T}_2(\theta, \phi) = e^{-i\frac{\theta}{2}} T_2(\theta, \phi)
\end{equation}
where we use a single phase shift $\theta$ instead of a differential phase shift in the internal phase shift of the MZI. The fundamental function of the MZI is to be able ``nullify'' (minimize to zero) power in either of its output powers given any input vector. In mathematical terms, given any $\bm{x}$, we should be able to generate an output of the form $\bm{y} = (y_1, 0)$. As defined in Refs. \citenum{Pai2020ParallelNetwork, Miller2020AnalyzingNetworks}, we can perform the nullification of $y_2$ for any MZI $T_2(\theta, \phi)$ with inputs $x_1, x_2$:
\begin{equation} \label{eqn:nullify}
    \begin{aligned}
        \theta &= 2\arctan\left|\frac{x_1}{x_2}\right| \\
        \phi &:= -\arg\left(\frac{x_1}{x_2}\right),
    \end{aligned}
\end{equation}
with the convention for $\theta, \phi$ being internal and external phase shifters as defined in Fig. \ref{fig:backprop} of the main text.

\subsection{Vector units}

\begin{figure*}
    \centering
    \includegraphics[width=\textwidth]{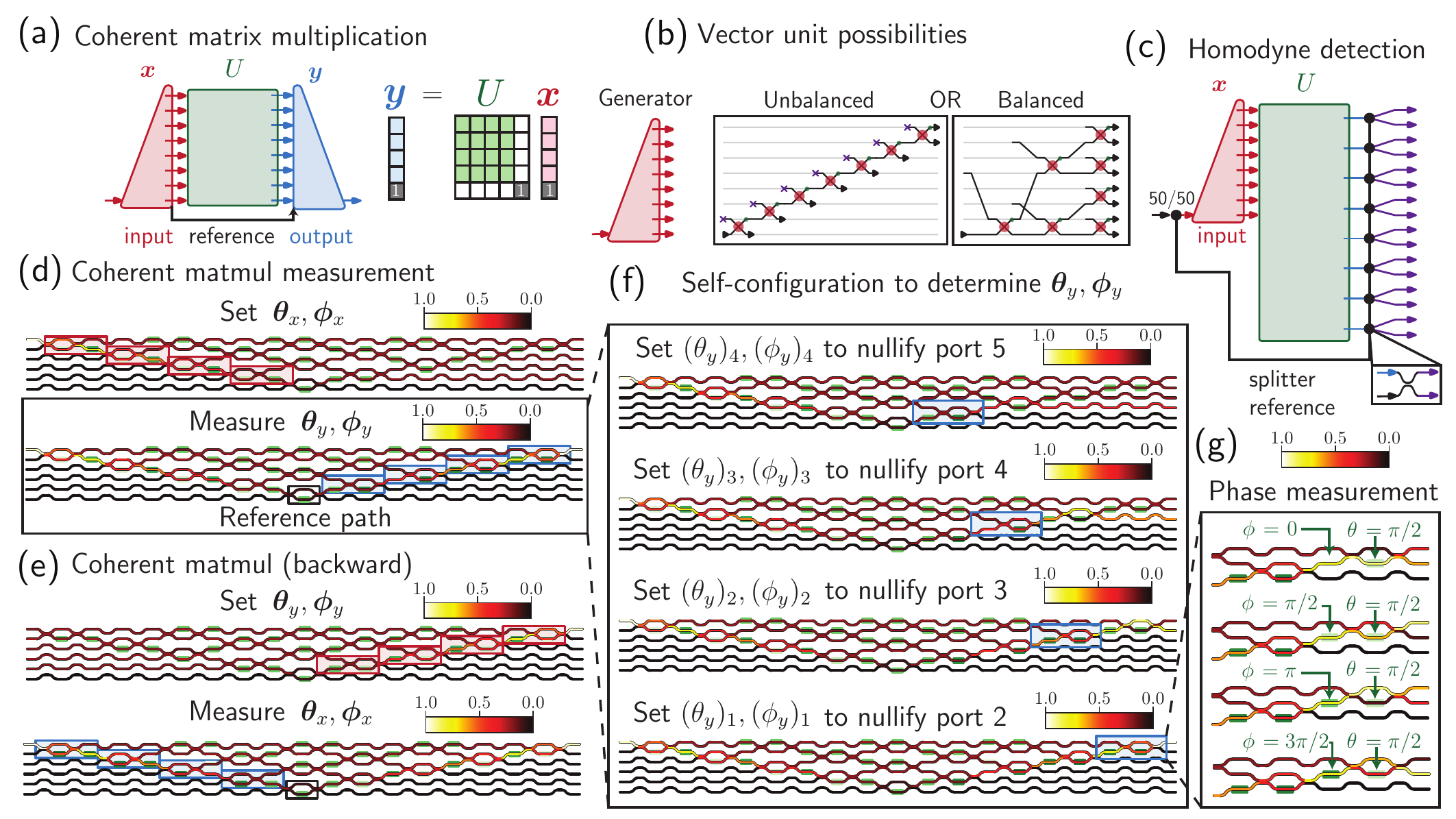}
    \caption{(a) In our $6 \times 6$ MZI network, forward coherent matrix $4 \times 4$ multiplication is performed using a generator and analyzer (input/output vector unit) configuration and a fifth reference dimension. (b) A vector unit can be unbalanced or balanced \cite{Miller2020AnalyzingNetworks}. (c) An alternative and likely faster approach is to use coherent detection or homodyne detection to measure amplitudes and phases. (d) Coherent matmul operation is performed by sending in an input based on the calibrated phase shifts and then performing self-configuration on the output fields including a reference path dimension, and (e) backward coherent matrix multiplication follows the same procedure but backwards. These plots are derived from actual measurements performed on the device and inset colorbars represent powers in the device corresponding to the colored waveguides shown. (f) Self-configuration proceeds by nullifying ports 5 through 2 in descending order. (g) Nullification is achieved using phase measurement rather than analog feedback minimization as in \cite{Annoni2017UnscramblingModes} as this is more efficient in our device configuration.}
    \label{fig:phase}
\end{figure*}

As proposed in Ref. \citenum{Miller2020AnalyzingNetworks}, a vector unit is a may be used as a $1 \times N$ input vector generator or an $N \times 1$ output vector analyzer (the flipped version of the input generator). An $N$-vector unit is a ``tree network'' of $N - 1$ splitters $\bm{\theta}_X$ and $N$ output phases $\bm{\phi}_X$, with the fully balanced binary tree and the maximally unbalanced linear cascade (diagonal line) as extreme cases (see the red and blue structures of Fig. \ref{fig:phase}(b)). An input generator generates optical modes representing any $N$-dimensional complex vector given a single input to the system up to a (nonphysical) global phase and can be either balanced or unbalanced as shown in Fig. \ref{fig:phase}(c). Operated in reverse, the analyzer allows for the $N - 1$ splitters to route all input light into a single port.

The overall mathematics can be represented in either vector or bra-ket notation as follows (where $X^\dagger$ represents analysis and $X$ represents generation):
\begin{equation}
    \begin{aligned}
        \bm{x} &= X \bm{e}_1\\
        |\bm{x} \rangle &= X |0\rangle\\
        X^\dagger |\bm{x} \rangle &= |0\rangle
    \end{aligned}
\end{equation}

The algorithm required to set up an \textit{output} vector unit analyzer requires first establishing a path between the root MZI and all other vector unit MZIs (known as a topological order) such that all the light exits the output of the vector unit Refs. \citenum{Pai2020ParallelNetwork, Miller2020AnalyzingNetworks}. 

Experimentally, as proven in Refs. \citenum{Pai2020ParallelNetwork, Miller2013Self-configuringInvited, Miller2020AnalyzingNetworks}, this can be achieved using self-configuration by minimizing the power (first sweeping $\phi$ and then sweeping $\theta$) for $N - 1$ open ports of the device to maximize output port power. All devices belonging to a given column can be programmed simultaneously (in parallel), so for binary tree architectures, this measurement can be done in $O(\log N)$ steps.

In this work, however, since we use a camera for all photodetection measurements, this protocol can be relatively slow. Therefore, we instead use four measurements, with $\theta = \pi / 2$ and $\phi = 0, \pi / 2, \pi, 3 \pi / 2$ to deduce the powers $p_0, p_{\pi / 2}, p_{\pi}, p_{3\pi / 2}$ and compute relative phase as $\arctan\left(\frac{p_{3\pi / 2} - p_{\pi / 2}}{p_{\pi} - p_{0}}\right)$. This is shown in Fig. \ref{fig:phase}(g).

Output detection can be made faster if necessary using homodyne coherent detection as shown in Fig. \ref{fig:phase}(c) where nominally 50 percent of the total input light is split into $N$ waveguides and sent directly to the output of the matrix unit implementing $U$. In the analog domain this protocol requires only a single step. As with our self-configuration phase measurement protocol, it requires additional computation on the digital end to deduce the phase.

\subsection{Matrix unit}

The matrix unit, shown in green in Fig. \ref{fig:backprop} in the main text, is any suitable arrangement of interferometers needed to represent a subset of unitary matrices in $\mathrm{U}(N)$; a \textit{universal} (unitary) matrix unit can implement any unitary matrix in $\mathrm{U}(N)$. Examples of universal matrix units are triangular \cite{Reck1994ExperimentalOperator, Miller2013Self-configuringInvited}, rectangular \cite{Clements2016AnInterferometers}, cascaded binary tree \cite{Miller2013Self-aligningCoupler}, and cosine-sine decomposition \cite{Mottonen2004QuantumGates} (which is more useful for quantum applications of this scheme, but can be represented classically). Such devices consist of $O(N^2)$ parameters: $N(N - 1) / 2$ MZIs with 2 phase shifters each $\bm{\theta}_U, \bm{\phi}_U$ and $N$ output phase shifters $\bm{\gamma}_U$.

Because multiplying by $\bm{\gamma}_U$ is an $O(N)$ operation, all computation for $\bm{\gamma}_U$ (both forward and backward passes in the gradient computation) is performed on the computer. In the protocol shown in Fig. \ref{fig:phase} and in the main text, we do not include any $\bm{\gamma}_U$ phase shifts due to the assumption that those computations are relatively inexpensive and can be fully accounted for off-chip.

The matrix unit is represented by an operator $U$ that performs the following operation (in vector notation and bra-ket notation):
\begin{equation}
    \begin{aligned}
        U \bm{x} &= \bm{y} \\
        U | \bm{x} \rangle &= |\bm{y} \rangle = Y |0\rangle
    \end{aligned}
\end{equation}
In vector notation, the relative phases given by $\arg (\frac{U \bm{x}}{\bm{x}})$ can be measured only up to an overall phase, so an additional measurement is required to measure this overall phase. In bra-ket notation, we typically can only ensure $\langle 0|Y^\dagger U X|0\rangle = e^{i \phi_0}$, where $\phi_0$ is some phase that depends on the effective overall path length in the device, which is a function of all the phase shifts. In theory, we could figure out what this overall path length is by some $O(N^2)$ mathematical computation, but in practice, this can be measured directly in $O(1)$.

\subsection{Reference arm}

Phase shifts in physical systems typically have no meaning without a reference, and this is ultimately crucial for designing and programming a photonic mesh. Adding a reference arm waveguide to an $N$-waveguide photonic mesh (mathematically, embedding all $N$-dimensional Hilbert space operations in an $N + 1$-dimensional Hilbert space), an example of which has previously been demonstrated in coherent detection for complex optical neural networks \cite{Zhang2021AnNetwork}.

Independent of reference arm placement, we treat the unitary operator ($U$ embedded in $N + 1$-dimensional Hilbert space as shown in Fig. \ref{fig:phase}(d) for $N = 4$) as follows:
\begin{equation} \label{eqn:refphase}
    \begin{bNiceArray}{C}[margin]\Block{3-1}{\bm{y}} \\ \\ \\ \\ \hline $z$ \end{bNiceArray} = \begin{bNiceArray}{CCC|C}[margin]\Block{3-3}{U} & & & 0 \\& \hspace*{1cm} & & $\vdots$ \\& & & $0$ \\\hline0 & $\hdots$& 0 & 1\end{bNiceArray} \begin{bNiceArray}{C}[margin]\Block{3-1}{\bm{x}} \\ \\ \\ \\ \hline $z$\end{bNiceArray},
\end{equation}
which allows us to calculate all phases in the matrix-vector multiplication relative to the phase shift in the added spatial mode (reference waveguide path length). We now can program and/or measure the full input and output $\bm{x}, \bm{y}$ no matter what settings are used for $U$. Assuming a total power of 1, the constant phasor $z$ here denotes a constant amplitude, such as $1 / \sqrt{N + 1}$ or whatever is deemed sufficient.

To properly measure phases for an $N \times N$ operation, we set the phase for the ($N + 1$)th output of any vector unit as the ``reference phase arm'' (shown throughout Fig. \ref{fig:phase}) and connect the reference arm to the waveguide where this phase is defined. If the magnitude of the $N$th element is zero, we choose that the reference phase of the vector is also zero. After storing the calibration curve of this reference phase in the computer, we can always set or measure this reference phase by maximizing power output of the reference arm MZI on the appropriate side of the device (e.g. as in the first step of Fig. \ref{fig:phase}(f)). This is generally a standard technique in phase detection in photonic circuits and similar schemes have been previously explored \cite{Zhang2021AnNetwork}.

Note that in the case of homodyne detection of Fig. \ref{fig:phase}(c), the math of the phase measurement is a bit different. A separate reference path is still provided, but instead of an analyzer with an additional reference dimension, the reference path is split and interfered at each output to determine the phase. This is a potentially faster and more ``standard'' method for measuring phases but the circuitry for bidirectional operation is a bit more complex.

\subsection{Calibration}

\begin{figure*}
    \centering
    \includegraphics[width=\textwidth]{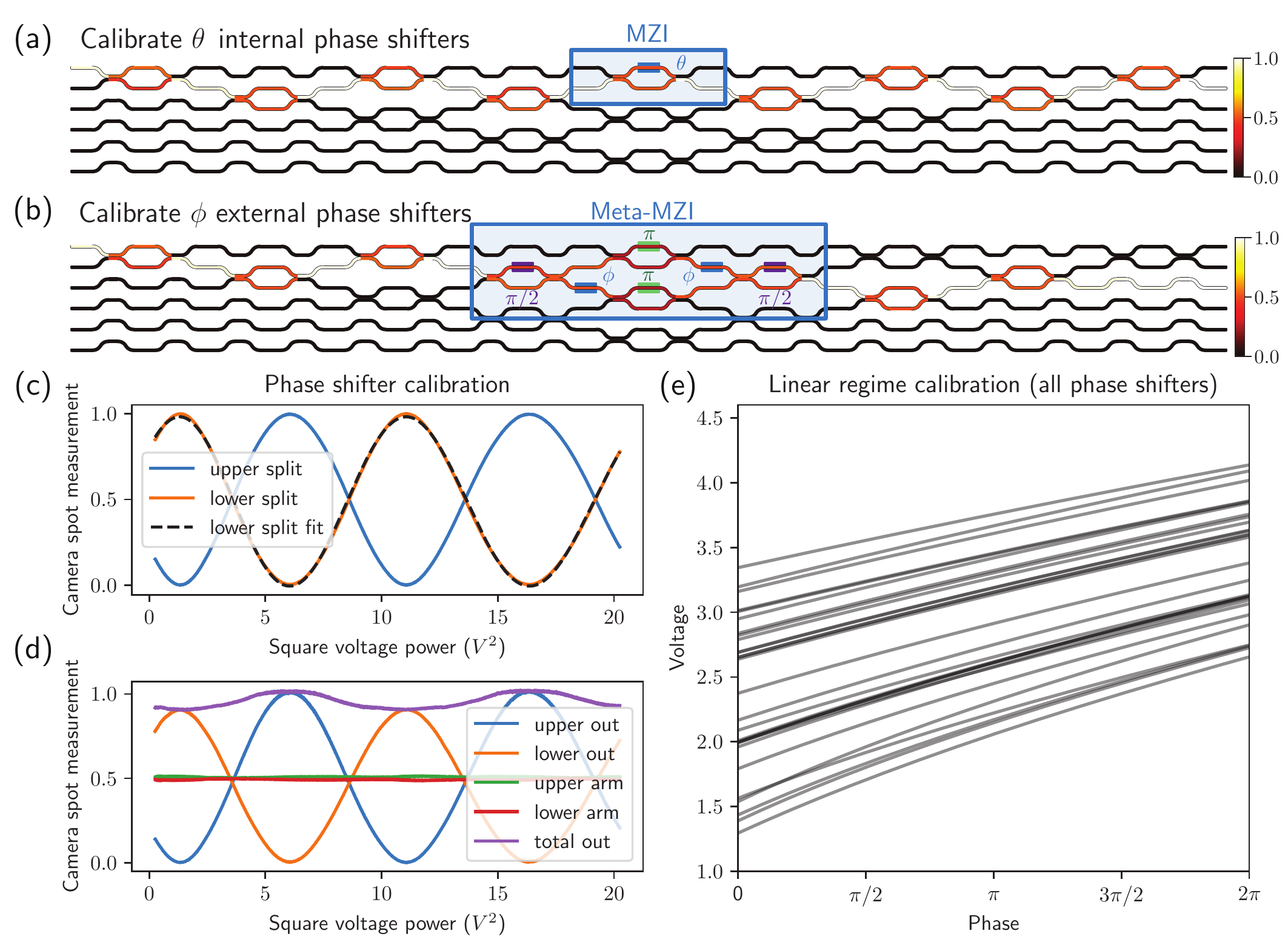}
    \caption{(a) Calibration of $\theta$ internal phase shifts using lightwires leading to MZIs. (b) Calibration of $\phi$ phase shifts using lightwires leading to meta-MZI structures created out of four neighboring MZIs. (c) Phase shifter calibration protocol shows excellent fit (d) Raw camera spot measurement shows that different grating taps have different coupling efficiencies (a source of error in gradient measurements). (e) Linear regime of the calibration curve shows the range of voltages that need to be applied to our phase shifters to ensure that a full $[0, 2\pi)$ range can be achieved.}
    \label{fig:calibration}
\end{figure*}

An important protocol for both vector units and matrix units is calibration of phase shifts for accurate inference and phase measurement. For our calibration protocol, we sweep phase shifts while recording an MZI split ratio measured using camera spots immediately after an assigned MZI depending on the calibrated phase shifter. An MZI split ratio can be represented in terms of a transmissivity $t = \sin^2 \theta$, where $\theta$ is twice the phase shift in the internal arm, which is used for calibration:
\begin{equation}
    t = \frac{p_t}{p} \approx \frac{p_t}{p_r + p_t}
\end{equation}
where $t$ is the transmissivity, $p$ is the total power at the input, $p_t$ is the cross state grating power and $p_r$ is the bar state grating power determined by summing up pixel values from the camera.

The model is:
\begin{equation}
    \begin{aligned}
        \theta &= p_0v^3 + p_1 v^2 + p_2 v + p_3\\
        t &= a \sin \theta + b.
    \end{aligned}
\end{equation}

Empirically, it suffices to fit $v^2 = q_0 \theta^3 + q_1 \theta^2 + q_2 \theta + q_3$ to convert voltage to phase.

For algorithmically calibrating the phase shifts, we use interferometers within the mesh to first calibrate all $\theta$ internal phase shifts from left to right by routing light via ``lightwires'' to all MZIs in the device, as shown in Fig. \ref{fig:calibration}(a) \cite{Mower2015High-fidelityCircuits}.

We then use ``meta-MZI'' structures within the mesh to calibrate all of the $\phi$ external phase shifters as shown in Fig. \ref{fig:calibration}(b). For this calibration, after we calibrate each of the $\phi$ phase shifters, we set $\phi = 0$ so that the other $\phi$ phase shifter in the meta-MZI has a consistent calibrated phase. Repeating this procedure for all $\phi$ phase shifts is sufficient to ensure that phase calibrations are all mutually consistent \cite{Mower2015High-fidelityCircuits}.

\begin{algorithm}[H]
    \caption{\textsc{Vector Unit Phase Conversion}}
    \label{alg:vecphase}
    \begin{algorithmic}[1]
        \Function{Vec2Phase}{$\bm{x}$} \Comment{Fig. \ref{fig:phase}(f)}
        \State \textbf{require} $\bm{x} \in \mathbb{C}^{N}, \|\bm{x}\| = 1$.
        \For{$m \in [1, 2, \ldots N - 1]$}
            \State $\phi_m \gets -\arg\left(\frac{x_1}{x_2}\right)$ \Comment{Fig. \ref{fig:phase}(g)}
            \State $\theta_m \gets 2\arctan\left|\frac{x_1}{x_2}\right|$ \Comment{nullify at $m + 1$}
            \State $x_m \gets e^{i\phi_m}\sin \frac{\theta_m}{2} x_m + \cos \frac{\theta_m}{2} x_{m + 1}$
            \State $x_{m + 1} \gets 0$
        \EndFor
        \State \Return $\bm{\theta}, \bm{\phi}$
        \EndFunction
        \item[]
        \Function{Phase2Vec}{$\bm{\theta}, \bm{\phi}$}
        \State \textbf{require} $\bm{\theta} \in [0, \pi]^{N}$.
        \State \textbf{require} $\bm{\phi} \in [0, 2\pi)^{N}$.
        \State $\bm{x} = [1, 0, \ldots 0] \in \mathbb{C}^N$
        \For{$m \in [1, 2, \ldots N - 1]$}
            \State $\begin{pmatrix} x_m \\ x_{m + 1}\end{pmatrix} \gets \widetilde{T}_2(\theta_m, \phi_m)^T \begin{pmatrix} x_m \\ x_{m + 1}\end{pmatrix}$
        \EndFor
        \State \Return $\bm{x} \exp(-i\arg(x_N))$ \Comment{Zero phase for $x_N$} 
        \EndFunction
    \end{algorithmic}
\end{algorithm}

\begin{algorithm}[H]
    \caption{\textsc{Forward Step}}
    \label{alg:forward}
    \begin{algorithmic}[1]
        \Function{MeshForward}{$\bm{x}$; $\bm{\theta}$, $\bm{\phi}$, $\bm{\gamma}$}
        \State \textbf{require} $\bm{x} \in \mathbb{C}^{N}, \|\bm{x}\| = 1$.
        \State \textbf{require} $\bm{\theta} \in [0, \pi]^{N(N - 1) / 2}$.
        \State \textbf{require} $\bm{\phi} \in [0, 2\pi)^{N(N - 1) / 2}$.
        \State \textbf{require} $\bm{\gamma} \in [0, 2\pi)^{N}$.
        \State $\bm{x} \gets [\bm{x} \cdot \sqrt{1 - 1 / N}, \sqrt{1 / N}]$ \Comment{add reference path}
        \State $\bm{\theta}_X, \bm{\phi}_X = \textsc{Vec2Phase}(\bm{x})$ \Comment{off-chip}
        \State $i \gets 1$
        \State $\bm{p} \gets \bm{0}$
        \State $\bm{w} \gets \textsc{SendForward}(\bm{\theta}_X, \bm{\phi}_X)$ \Comment{on-chip}
            \For{$n \in [1, 2, \ldots N - 1]$} \Comment{on-chip}
            \For{$m \in [1, 2, \ldots N - m]$}
                \State $\begin{pmatrix} w_m \\ w_{m + 1}\end{pmatrix} \gets \widetilde{T}_2(\theta_i, \phi_i) \begin{pmatrix} w_m \\ w_{m + 1}\end{pmatrix}$ \Comment{forward prop}
                \State \textbf{measure} $p_{\theta_i}, p_{\phi_i}$ \Comment{detect phase shift powers}
                \State $i \gets i + 1$
            \EndFor
            \EndFor
            \State $\bm{\theta}_Y, \bm{\phi}_Y = \textsc{ReadForward}(\bm{w})$ \Comment{self-configuration}
            \State $\bm{y} \gets \textsc{Phase2Vec}(\bm{\theta}_Y, \bm{\phi}_Y) \cdot e^{i\bm{\gamma}}$ \Comment{off-chip}
            \State $\bm{y} \gets \bm{y}_{:N} / \sqrt{1 - 1 / N}$ \Comment{remove reference}
            \State \Return $\bm{y}, \bm{p}$
        \EndFunction
    \end{algorithmic}
\end{algorithm}

\begin{algorithm}[H]
    \caption{\textsc{Backward Step}}
    \label{alg:backward}
    \begin{algorithmic}[1]
        \Function{MeshBackward}{$\bm{y}$; $\bm{\theta}$, $\bm{\phi}$, $\bm{\gamma}$}
        \State \textbf{require} $\bm{y} \in \mathbb{C}^{N}, \|\bm{y}\| = 1$.
        \State \textbf{require} $\bm{\theta} \in [0, \pi]^{N(N - 1) / 2}$.
        \State \textbf{require} $\bm{\phi} \in [0, 2\pi)^{N(N - 1) / 2}$.
        \State \textbf{require} $\bm{\gamma} \in [0, 2\pi)^{N}$.
        \State $\bm{y} \gets [\bm{y} \cdot \sqrt{1 - 1 / N}, \sqrt{1 / N}]$ \Comment{add reference path}
        \State $\bm{\theta}_Y, \bm{\phi}_Y = \textsc{Vec2Phase}(\bm{y}^* \cdot e^{i\bm{\gamma}})$ \Comment{off-chip}
        \State $i \gets N(N - 1) / 2$
        \State $\bm{p} \gets \bm{0}$
        \State $\bm{w} \gets \textsc{SendBackward}(\bm{\theta}_Y, \bm{\phi}_Y)$ \Comment{on-chip} 
        \For{$n \in [1, 2, \ldots N - 1]$} \Comment{on-chip}
            \For{$m \in [1, \ldots m]$}
                \State $\begin{pmatrix} w_m \\ w_{m + 1}\end{pmatrix} \gets \widetilde{T}_2(\theta_i, \phi_i)^T \begin{pmatrix} w_m \\ w_{m + 1}\end{pmatrix}$ \Comment{back prop}
                \State \textbf{measure} $p_{\theta_i}, p_{\phi_i}$ \Comment{detect phase shift powers}
                \State $i \gets i - 1$
            \EndFor
        \EndFor
        \State $\bm{\theta}_X, \bm{\phi}_X = \textsc{ReadBackward}(\bm{w})$ \Comment{self-configuration}
        \State $\bm{x} \gets \textsc{Phase2Vec}(\bm{\theta}_X, \bm{\phi}_X)$ \Comment{off-chip}
        \State $\bm{x} \gets \bm{x}_{:N} / \sqrt{1 - 1 / N}$ \Comment{remove reference}
        \State \Return $\bm{x}, \bm{p}$
        \EndFunction
    \end{algorithmic}
\end{algorithm}

\begin{algorithm}[H]
    \caption{\textsc{In Situ Backpropagation}}
    \label{alg:backprop}
    \begin{algorithmic}[1]
        \Function{InSituGradient}{$\bm{x}$, $\bm{z}$, $\ell$}
        \State \textbf{require} $\bm{x} \in \mathbb{C}^{N}$.
        \State \textbf{require} $\bm{z} \in \mathbb{R}^{N}$.
        \State $\bm{\theta}, \bm{\phi}, \bm{\gamma} \gets \textsc{Phases}(U^{(\ell)})$ \Comment{phases of mesh $\ell$}
        \State $P \gets \|\bm{x}\|^2$ \Comment{input power scaling}
        \State $\bm{y}, \bm{p} \gets \textsc{MeshForward}(\bm{x} / \sqrt{P}, \bm{\theta}, \bm{\phi}, \bm{\gamma})$
        \If{$\ell = L$}\Comment{End of the neural net}
            \State $\bm{y}_{\mathrm{aj}} = \partial c(\hat{\bm{z}}, \bm{z}) / \partial \bm{z}$ \Comment{or $\partial \mathcal{L} / \partial \bm{z} |_{\bm{x}, \bm{z}}$}
            \State $\bm{g}_{\mathrm{tot}} = \emptyset$ \Comment{Empty gradient set}
        \Else
            \State $\bm{x}_{\mathrm{aj}}, \bm{g}_{\mathrm{tot}} \gets \textsc{InSituGradient}(f_\ell(\bm{y}), \bm{z}, \ell + 1)$
            \State $f_{\mathrm{vjp}}^{(\ell)} \gets \textsc{VJP}(f^{(\ell)})$ \Comment{Autodiff, JAX/Haiku}
            \State $\bm{y}_{\mathrm{aj}} \gets f_{\mathrm{vjp}}^{(\ell)}(\bm{y}, \bm{x}_{\mathrm{aj}})$
        \EndIf
        \State $P_{\mathrm{aj}} \gets \|\bm{y}_{\mathrm{aj}}\|^2$ \Comment{adjoint power scaling}
        \State $\bm{x}_{\mathrm{aj}}, \bm{p}^{\mathrm{aj}} \gets \textsc{MeshBackward}(\bm{y}_{\mathrm{aj}} / \sqrt{P_{\mathrm{aj}} }, \bm{\theta}, \bm{\phi}, \bm{\gamma})$
        \State $\ldots, \bm{p}^{\mathrm{sum}} \gets \textsc{MeshForward}(\bm{x} - i\bm{x}_{\mathrm{aj}}^*, \bm{\theta}, \bm{\phi}, \bm{\gamma})$
        \State $\bm{g} \gets \bm{p}^{\mathrm{sum}} - \bm{p}^{\mathrm{aj}} - \bm{p}$ \Comment{analog or digital optical VJP}
        \State $\bm{g} \gets \bm{g} \cdot \sqrt{PP_{\mathrm{aj}}} / 2$ \Comment{scaling factor}
        \State \Return $\bm{x}_{\mathrm{aj}}, [\bm{g}, \bm{g}_{\mathrm{tot}}]$ \Comment{Append new gradients $\bm{g}$ to $\bm{g}_{\mathrm{tot}}$}
        \EndFunction
    \end{algorithmic}
\end{algorithm}

\begin{algorithm}[H]
    \caption{\textsc{In Situ Backpropagation Training}}
    \label{alg:backproptrain}
    \begin{algorithmic}[1]
        \Function{InSituMinibatchTrain}{$X$, $Z$, $B$}
        \State \textbf{require} $X \in \mathbb{C}^{N_{\mathrm{train}} \times N}$.
        \State \textbf{require} $Z \in \mathbb{R}^{N_{\mathrm{train}} \times N_{\mathrm{label}}}$.
        \State $\bm{h} \gets \bm{0}$ \Comment{tracks gradient history}
        \For{$t \in [1, 2, \ldots T]$} \Comment{on-chip}
            \State randomly sample $X_{t}, Z_{t}$ from $X, Z$.
            \State \textbf{require} $X_t \in \mathbb{C}^{B \times N}$.
            \State \textbf{require} $Z_t \in \mathbb{R}^{B \times N_{\mathrm{label}}}$.
            \For{$\bm{x}_b, \bm{z}_b \in X_{t}, Z_{t}$}
                \State $\cdots, \bm{g}_b \gets \textsc{InSituGradient}(\bm{x}_b, \bm{z}_b, \ell + 1)$
            \EndFor
            \State $\bm{g}_t \gets \sum_{b = 1}^B \bm{g}_b / B$ \Comment{minibatch average}
            \State $\delta\bm{\eta}, \bm{h} \gets$ \textsc{Optimizer}$(\bm{g}_t, \bm{h}$)
        \EndFor
        \EndFunction
    \end{algorithmic}
\end{algorithm}

\section{Pseudocode}

In this section, we specifically provide some pseudocode required to implement various algorithms required for \textit{in situ} backpropagation on our triangular mesh platform. Note that these approaches can be implemented on any matrix unit provided that the vector units can be used to generate any input fields. For output field generation, one can self-configure for backward and forward measurements on the existing vector units (Fig. \ref{fig:phase}(f)) or use a homodyne vector unit for measurement (Fig. \ref{fig:phase}(g)).

Our recursively defined algorithm for backpropagation on photonic meshes using the call $\bm{g} = \textsc{InSituGradient}(\bm{x}, \bm{z}, 1$) where $\bm{g}$ here represents gradients taken over all $\bm{\eta}$ in the network, as defined in Alg. \ref{alg:backprop}, is based on Algs. \ref{alg:vecphase}, \ref{alg:forward}, \ref{alg:backward} for generator/analyzer operation and the forward/backward steps for backpropagation. Note that some of the procedures such as \textsc{ReadBackward}, \textsc{SendBackward}, \textsc{ReadForward}, \textsc{SendForward}, \textsc{Phases} do not have pseudocode, but these are explained in our Methods section and in Refs. \citenum{Miller2015PerfectComponents, Miller2020AnalyzingNetworks}.

As previously discussed (Methods), the \textsc{VJP} (or vector Jacobian product) function is often used in neural networks and autodifferentiation frameworks (e.g., JAX) to automatically carry out chain rule steps used in measuring gradients. As defined in Alg. \ref{alg:backprop}, a VJP calculation based on nonlinearity derivatives is performed in the digital domain since the nonlinearity itself is also performed in the digital domain. We have already defined VJP in the context of optical backpropagation (``optical VJP'') in the Methods section in terms of physical measurement; in general, nonlinear VJPs are more straightforward to compute digitally. Computing nonlinear VJPs does not offer much benefit in the optical domain for our purpose (energy efficient computation) since the energy to define inputs and outputs is already $O(N)$ in the digital-analog conversion which is also the complexity of an elementwise digital nonlinearity.

Finally, now that we have defined all of the gradient measurement pseudocode, we are ready to define the final training protocol, which we use throughout this paper to achieve photonic \textit{in situ} training. We define the full training set of $N_{\mathrm{train}}$ training examples as a $N_{\mathrm{train}} \times N$ data matrix $X$ and associated label set $N_{\mathrm{train}} \times N_{\mathrm{label}}$ $Z$:

Note that there are two nontrivial implementations in Alg. \ref{alg:backproptrain}: the Adam optimizer \cite{Kingma2015Adam:Optimization} and minibatch training protocols. In practice, we leverage autodifferentiation packages to implement much of this needed functionality (e.g., we use JAX's optim package for the Adam optimizer). We choose a minibatch size of 1 implementing a purely ``stochastic'' update which does not average over many training examples. This helps to avoid errors in the gradient which as we have found can accumulate over a large batch of training examples. This further underscores the importance of reducing gradient error to enable minibatch training.

Additionally, further investigation is warranted to explore \textit{analog} adaptive update schemes that store previous gradients in nonvolatile memory. This would be important in cases where a purely analog update is required; otherwise a potentially more energy-consuming digital subtraction update would be needed to compute the history aggregation vector $\bm{h}$ at each step of the optimization.

\section{Analog update}

\subsection{Equivalence of digital and analog update}
Here, we prove the equivalence of $d_\eta(0)$ (our new analog measurement proposal) and the numerically evaluated gradient $\mathcal{I}(x_\eta x_{\eta, \mathrm{aj}})$ which has been shown to be equivalent to the digital subtraction update \cite{Hughes2018TrainingMeasurement}.

The idea is to input a varying sum signal $\bm{x} - i \bm{x}^*_\mathrm{aj} e^{i\zeta}$ and analyze the varying or AC component $d_\eta(\zeta)$ of the power $p_\eta(\zeta)$ measured at phase shifter $\eta$ which has the fields $x_\eta$ when sending $\bm{x}$ alone and $ix^*_{\mathrm{aj}, \eta}e^{i\zeta}$ when sending $\bm{x}^*_\mathrm{aj}$ alone:
\begin{equation}
\begin{aligned}
    p_\eta(\zeta) &= |x_\eta|^2 + |ix^*_{\mathrm{aj}, \eta}e^{i \zeta}|^2 -  2\mathcal{R}(ix_\eta x^*_{\mathrm{aj}, \eta}e^{i \zeta}) \\
    d_\eta(\zeta) &= -2\mathcal{R}(ix_\eta x^*_{\mathrm{aj}, \eta}e^{i \zeta}) = 2\mathcal{I}(x_\eta x_{\mathrm{aj}, \eta}e^{i \zeta}),
\end{aligned}
\end{equation}
which is equivalent to the gradient iff $\zeta = 0$.

\begin{figure*}
    \centering
    \includegraphics[width=\textwidth]{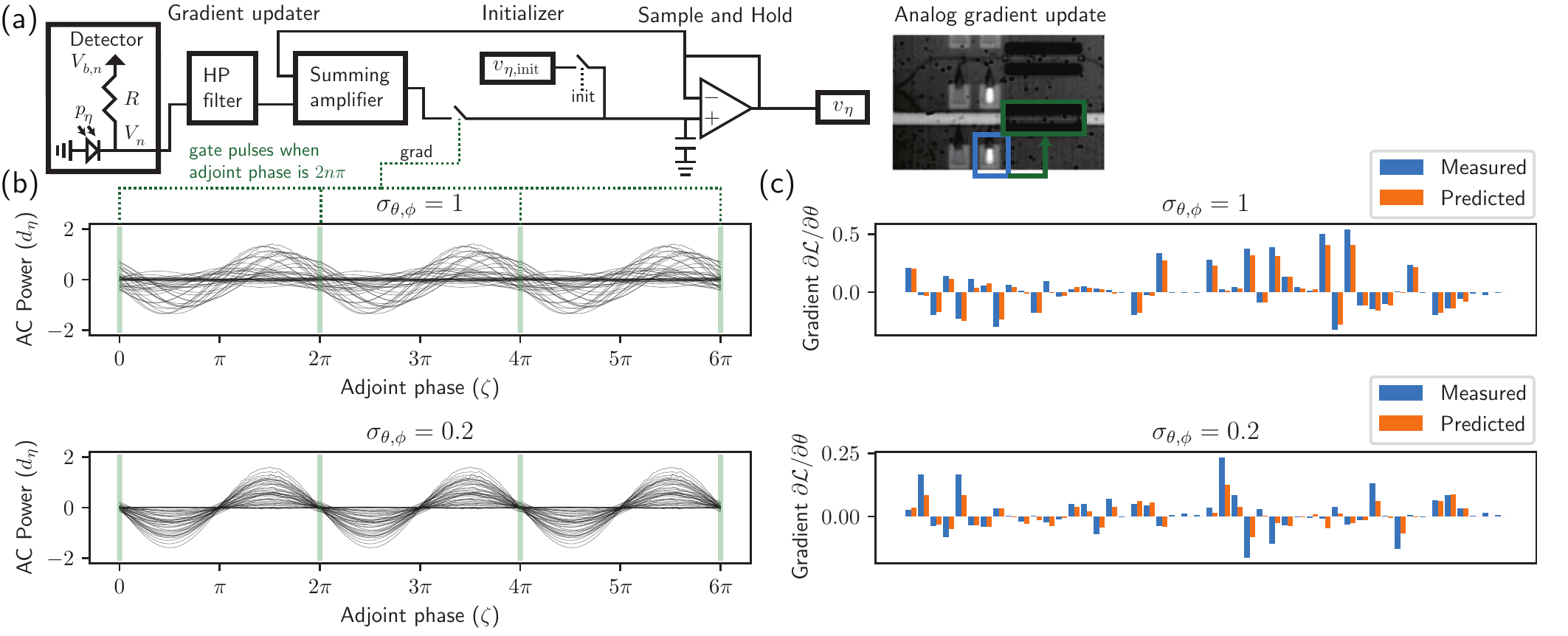}
    \caption{(a) Our conceptual analog gradient update flow for updating phase shifter $\eta$ based on power signal $p_\eta(\zeta)$, which varies according to adjoint phase. An integrated detector is connected via a ``gradient updater'' circuit consisting of a high-pass (HP) filter and summing amplifier to a sample-and-hold circuit scheme with an initializer for direct setting of voltage. (b-c) Standard deviation in the phase shift error is used to specify either far from or close to convergence ($\sigma_{\theta, \phi} = 1, 0.2$ respectively). (b) The AC power signal $d_\eta$ versus the adjoint phase is experimentally measured on our chip across all relevant grating tap monitors, showing a decrease in gradient magnitude and more ``in-phase'' behavior near convergence. (c) As the distance to convergence decreases, there is more error in the gradient computation as expected, which is more explicitly shown in Fig. \ref{fig:backpropgradient}(f).}
    \label{fig:analog}
\end{figure*}

\subsection{Analog update protocol}

Now that we have shown the equivalence of the digital and analog updates, we discuss the physical implementation of the analog gradient update implementation in hybrid photonic neural networks. As discussed in the main text, the analog signal processing to implement the gradient updater involves (1) a high pass filter and (2) a gated integrator implemented using a summing amplifier feedback (with gate width specified in the original signal synchronized to $\zeta(t) = 2\pi n$). The output of the integrator is the gradient that can be directly applied as a control signal to the sample-and-hold phase shifter voltage. This is shown in more detail in Fig. \ref{fig:analog}(a).

Constant scaling factors required for gradient updates may be reflected in the analog signal processing, e.g. in the integrator step. Note that during \textit{in situ} backpropagation, the forward- and backward- propagating optical signals in each of the photonic mesh accelerator chips are normalized to the same power. The computer stores the actual vector norms of the input and output vectors $\bm{x}, \bm{x}_{\mathrm{aj}}$ as $P, P_{\mathrm{aj}}$ as defined also in Alg. \ref{alg:backproptrain}. The sum vector $\bm{x} - i \bm{x}_{\mathrm{aj}}^*$ is trickier to rescale. In this case, the input light is split equally into two input vectors implementing the normalized $\bm{x}, \bm{x}_{\mathrm{aj}}^*$ and then interfered to yield the (lossy) vector sum $(\bm{x} - i \bm{x}_{\mathrm{aj}}^*) / \sqrt{2}$ as shown in Fig. \ref{fig:backpropgradient}(b) of the main text. To recover the gradient, all that is needed is to multiply by the normalized factor $\sqrt{P P_{\mathrm{aj}}}$. This can be applied as a uniform scaling factor to all gradient updaters used to determine the gradient in the analog domain. This is the only scaling factor that varies according to the training example sent through the device; all other scaling factors can be grouped in with the overall learning rate of the system.

\section{Simulated error analysis}

\begin{figure*}
    \centering
    \includegraphics[width=\textwidth]{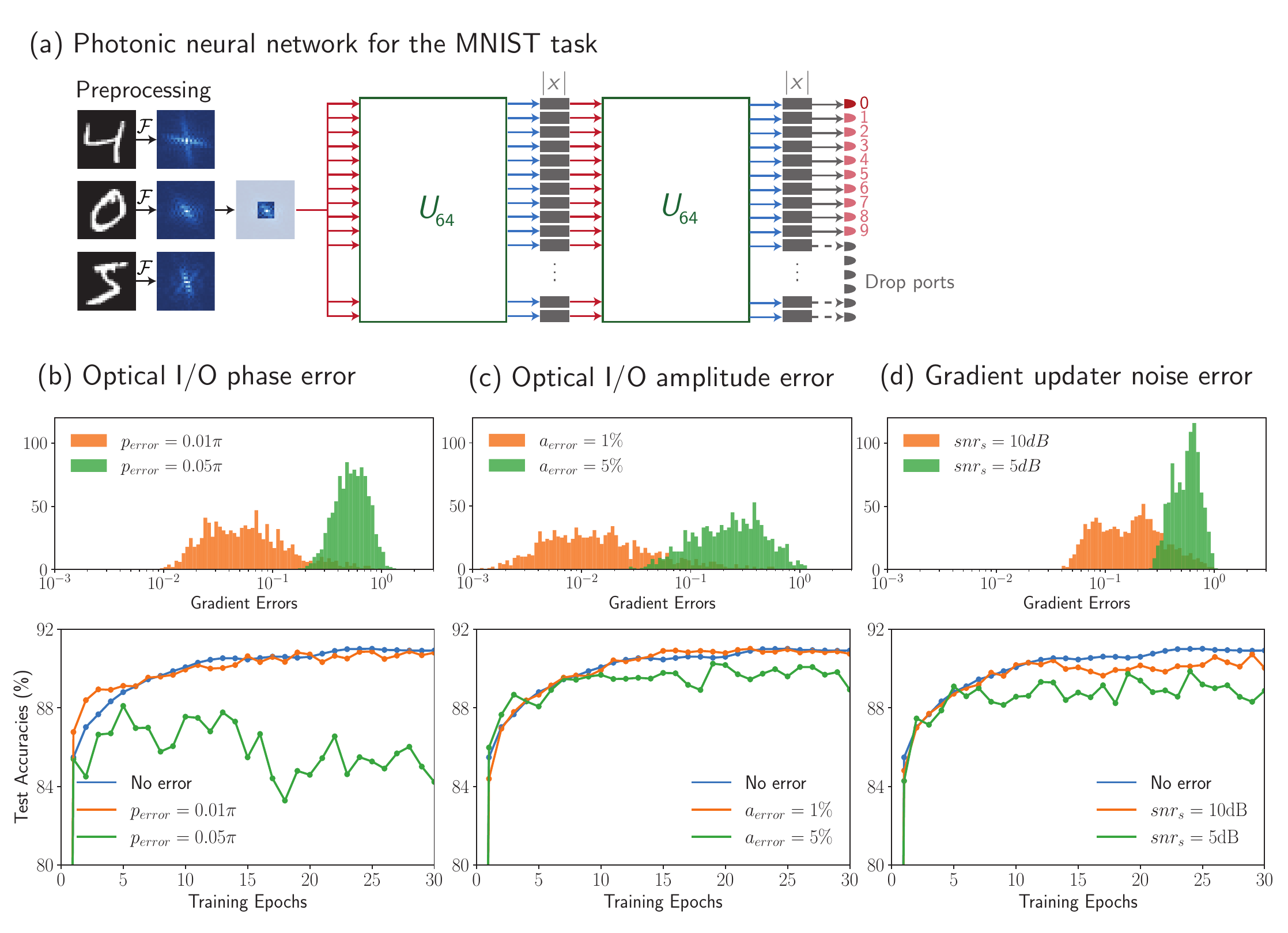}
    \caption{(a) Two-layer triangular mesh optical neural network with $N = 64$ inputs. Images from MNIST datasets are pre-processed following the procedure in~\cite{Pai2019ParallelNetwork, Williamson2020ReprogrammableNetworks}. (b)-(d) Normalized gradient errors (upper row) and testset accuracies (lower row) for models trained with in-situ backpropagation algorithm. Different types of hardware errors and noises are added to the training process; (b) field amplitude error $a_{error}$, (c) field error $p_{error}$, and (d) photon shot noise $s_{noise}$.}
    \label{fig:error_analysis}
\end{figure*}

The processing capability of our proposed experimental prototype is limited by the size of the photonic circuit since the circuit size is just $N = 4$. For completeness, we run simulations with a larger photonic circuit ($N = 64$) that uses a two-layer ``triangular mesh'' architecture with the same absolute value digital nonlinearity. We train this more expressive model on MNIST dataset~\cite{Deng2012TheResearch} for hand-written digits recognition. As shown in Fig.~\ref{fig:error_analysis}(a), we follow the pre-processing procedure in ~\cite{Pai2019ParallelNetwork} to convert the input $28\times28$ images into $64$-dimensional complex vectors that are then input into the photonic circuit. We use the Adam optimizer~\cite{Kingma2015Adam:Optimization} with learning rate $\alpha = 0.001$ to train the model following Alg. ~\ref{alg:backproptrain} for $30$ epochs. We use the digit classification accuracy on the testset (with unseen data samples) as the metric for model performance.

To evaluate the robustness of the in-situ backpropagation process with respect to hardware errors, we add three types of errors in the simulations:
\begin{enumerate}
    \item $a_{error}$, which represents the amplitude errors in field generation and analysis (\textsc{ReadForward}, \textsc{SendForward} in Alg.~\ref{alg:forward} and Alg.~\ref{alg:backward}).
    \item $p_{error}$, which represents phase errors in field generation and analysis.
    \item $s_{noise}$, which represents photon shot noise in optical power monitoring (line $14$ in Alg.~\ref{alg:forward} and Alg.~\ref{alg:backward}).
\end{enumerate}
We calculate the shot noise signal-noise-ratio (SNR) $snr_{s}$ at signal intensity $\sim 1/N$ since the input into optical neural network is normalized and each port has same average optical power. As shown in Fig.~\ref{fig:error_analysis}(b)-(d), with moderate level of noise (consistent with what is reported in current photonic circuits~\cite{Bandyopadhyay2021HardwarePhotonics}), the model convergence is minimally influenced, despite minor fluctuations. This demonstrates the robustness of in-situ backpropagation to noise and hardware errors, which are difficult to totally eliminate in modern analog computing systems. All data and code to reproduce these results are provided in our data availability repository \cite{Pai2022Solgaardlab/photonicbackprop:Networks}. 

\bibliography{training}

\end{document}